\documentclass[journal]{IEEEtran}
\usepackage{amsmath,epsfig,amssymb,verbatim,amsopn,subfigure,color}
\usepackage{cite,xspace}
\usepackage{array,algorithm,algorithmic}
\usepackage{bbm}
\usepackage{array}
\usepackage{multirow}
\usepackage{multicol}
\usepackage{rotating}
\usepackage{dsfont,float}
\usepackage{stfloats}
\usepackage{enumerate,makecell}
\usepackage{graphicx}
\usepackage{graphbox}
\usepackage{cases}
\usepackage{float}
\normalsize

\renewcommand{\baselinestretch}{1.05}

\makeatletter
\let\l@ENGLISH\l@english
\makeatother

\makeatletter
\renewcommand*{\@opargbegintheorem}[3]{\trivlist
  \item[\hskip \labelsep{\itshape #1\ #2}] {\itshape (#3):} {\normalfont}}
\makeatother
\usepackage{relsize}
\newcommand{\LL}  {\rm{L}}
\newcommand{\NN}  {\rm{N}}

\newtheorem{lemma}{Lemma}
\newtheorem{remark}{Remark}

\newtheorem{proposition}{Proposition}

\graphicspath{{figs/},{Figures/}}

\ifCLASSOPTIONpeerreview
\else
\fi
\newcommand{\AuthorOne}{Weihao~Wang}
\newcommand{\AuthorTwo}{Zesong~Fei, {\em{Senior Member, IEEE}}}
\newcommand{\AuthorThree}{Jing~Guo, {\em{Member, IEEE}}}
\newcommand{\AuthorFour}{Salman~Durrani, {\em{Senior Member, IEEE}}}
\newcommand{\AuthorFive}{Halim~Yanikomeroglu, {\em{Fellow, IEEE}}}

\newcommand{\ThankOne}{This work was supported in part by National Natural Science Foundation of China under Grant 62001029, and in part by Beijing Natural Science Foundation under Grant L202015. (Corresponding author: Jing Guo.)

W. Wang is with the School of Information and Electronics, Beijing Institute of Technology, Beijing 100081, China, and also with the Yangtze Delta Region Academy of Beijing Institute of Technology, Jiaxing 314000, China (Email: weihaowang@bit.edu.cn). Z. Fei and J. Guo are with the School of Information and Electronics, Beijing Institute of Technology, Beijing 100081, China (Emails: \{feizesong, jingguo\}@bit.edu.cn). S. Durrani is with the School of Engineering, The Australian National University, Canberra, ACT 2601, Australia (Email: salman.durrani@anu.edu.au). H. Yanikomeroglu is with the Non-Terrestrial Networks (NTN) Lab, Department of Systems and Computer Engineering, Carleton University, Ottawa, ON K1S 5B6, Canada (E-mail: halim@sce.carleton.ca).

Copyright (c) 20xx IEEE. Personal use of this material is  
permitted. However, permission to use this material for any other purposes must be obtained from the IEEE by sending a request to pubs-permissions@ieee.org.}

\begin{document}

\title{Outage Performance of Multi-tier UAV Communication with Random Beam Misalignment}
\author{\IEEEauthorblockN{\AuthorOne,~\AuthorTwo,~\AuthorThree,~\AuthorFour,~and~\AuthorFive\thanks{\ThankOne}}}
\maketitle

\vspace{+5mm}

\begin{abstract}
By exploiting the degree of freedom on the altitude, unmanned aerial vehicle (UAV) communication can provide ubiquitous communication for future wireless networks. In the case of concurrent transmission of multiple UAVs, the directional beamforming formed by multiple antennas is an effective way to reduce co-channel interference. However, factors such as airflow disturbance or estimation error for UAV communications can cause the occurrence of beam misalignment. In this paper, we investigate the system performance of a multi-tier UAV communication network with the consideration of unstable beam alignment. In particular, we propose a tractable random model to capture the impacts of beam misalignment in the 3D space. Based on this, by utilizing stochastic geometry, an analytical framework for obtaining the outage probability in the downlink of a multi-tier UAV communication network for the closest distance association scheme and the maximum average power association scheme is established. The accuracy of the analysis is verified by Monte-Carlo simulations. The results indicate that in the presence of random beam misalignment, the optimal number of UAV antennas needs to be adjusted to be relatively larger when the density of UAVs increases or the altitude of UAVs becomes higher.
\end{abstract}
%

\ifCLASSOPTIONpeerreview
    \newpage
\fi

\section{Introduction}\label{sec:intro}

\subsection{Background}
The emergence of novel applications, such as the Internet of Things, virtual/augmented reality, and telemedicine, has largely accelerated the growth of data and coverage demands nowadays. In order to satisfy the ever-increasing requirements of capacity and ubiquitous coverage, a promising way is the dense deployment of access points and base stations (BSs). However, for certain scenarios, e.g., the temporary hotspot regions or emergency areas, by considering the economic factors and the limitation of location, the deployment of terrestrial BSs may not be possible. Instead, an effective solution is to utilize unmanned aerial vehicles (UAVs) as the aerial BSs \cite{9383780,2019LiJIOT}. Compared to terrestrial infrastructures, due to the capability of mobility, UAVs can provide better link quality. Additionally, UAVs have the advantages such as rapid deployment and flexibility for reconfiguration, which highly extends the degree of freedom to provide ubiquitous coverage \cite{7470933,feng2018spectrum}. 
Because of the merits of UAV, it is anticipated that UAV-assisted wireless communication will be a critical enabling technology for beyond 5G and 6G networks \cite{8766143,9852974}. Note that, due to the flight regulations and hardware constraints, it is impractical for the UAVs flying at the same altitude, which necessitates the division of the network into multiple tiers \cite{8859647}. The multi-tier UAV
network, on the one hand, can balance the trade-off between line-of-sight (LoS) connectivity
and path loss, thereby improving the spectrum efficiency \cite{Hossain-2018}; on the other hand, aerial  BSs at different tiers can play distinct roles
\cite{7744808}, such as providing broad-area coverage or enhancing local-area communication capacity,
thereby offering a variety of communication services. Consequently, it is foreseen in the future that the applications of multi-tier UAV network
will become more widespread in the field of wireless communications \cite{9598918}.


\subsection{Motivation and Related Work}
Due to the limited spectrum resources, as the network becomes denser, co-channel interference is inevitable in a wireless communication system with multiple coexisting UAVs. As a result, a key research issue in the multi-tier UAV communication network is the performance degradation caused by severe interference \cite{8913615}. Among all possible approaches, by forming the directional beam pointing to the serving device, beamforming is a highly effective technology for reducing severe interference and improving the reliability of the communication system.

The UAV communication assisted by beamforming technology has received much attention. Therein, some works studied the robust beamforming design for UAV communications. For example, the authors in \cite{8907440} aimed to maximize the achievable sum rate in a multi-user UAV system by jointly optimizing the location of UAVs and the beamforming vector. In \cite{9311842}, a machine learning based channel estimation approach was developed to overcome the outdated channel state information, based on which a beamforming compensation algorithm was designed to reduce the interference. In \cite{9779715}, a precoding scheme was proposed for the high-altitude platform stations and terrestrial BSs coexisted network, which can generate high-gain beams in the direction of users and nulls in the direction of terrestrial BSs.  More relevant papers are surveyed in \cite{9598918}. Moreover, to better understand the impact of system parameters \cite{9378781,Haenggi-2012}, there are some works conducting the system-level performance evaluation for beamforming-assisted UAV communications based on stochastic geometry. Specifically, the paper \cite{8422376} analyzed the probability of successfully establishing a backhaul and the expected data rate, and only the influence of the main-lobe beam was considered. In \cite{9250029}, the authors analyzed the downlink coverage and rate for the aerial user with the incorporation of directional beamforming, where a finite 2D binomial point process was used to model the locations of aerial-BSs and two spectrum sharing policies were included. For a 3D UAV network, the authors in \cite{8856258} used a 3D blockage process and antenna pattern to capture the influence of UAV position when measuring the coverage probability in both the uplink and downlink. The work in \cite{9665243} investigated the coverage probability for a 3D distributed UAV communication network with the incorporation of zero-forcing beamforming and joint transmission. Note that the aforementioned works are considered an ideal case, i.e., the beam between the transmitter and receiver is perfectly aligned. However,  the perfect beam alignment may be impossible in practice. On the one hand, different from the ground infrastructure, UAVs are hovering or flying in the sky, and airflow disturbances or engine vibrations can cause the jitter of UAVs. Due to the movement and jitter, beam misalignment for UAV communications is unavoidable. On the other hand, the error estimation for the angle of arrival may occur, which will undoubtedly lead to imperfect beam alignment. Consequently, for beamforming-assisted UAV communication, it is necessary to take the influence of beam misalignment into account. 

Currently, there are a few papers that have investigated beamforming for UAV communications by considering imperfect beam alignment. More specifically, a prototype equipped with a steering mechanism to support the connection between a UAV and a BS was presented in \cite{8533583}, based on which the signal strength was measured with the consideration of unstable beam pointing. The authors in \cite{8633302} first analyzed the impact of beam deviation, and they proposed a method that leverages spatial angle quantification and the improved orthogonal matching pursuit algorithm to achieve better precoding performance. An approach to derive the channel capacity by considering the beam misalignment was presented in \cite{9614341}, where the classical control process was used to characterize the mobility of the UAV. The study in \cite{9707733} analyzed the performance of a free-space optical communication system with beam misalignment in 3D space, where a single UAV acted as a relay node to serve the terrestrial UE. In \cite{9214955}, via the utilization of both the navigation and channel information, the authors designed a robust beamforming algorithm, which can make the beam weight factor tolerate the direction of arrival error. 

Note that most of the aforementioned works only focused on the single transmitter, and the characterization of the interference in a large-scale network has been largely ignored. In the field of terrestrial wireless networks, based on stochastic geometry, the work in \cite{wildman2014joint} investigated the impact of beam misdirection on the coverage and throughput performance of a large-scale wireless network. In \cite{8345639}, the authors analyzed the coverage probability of the mmWave decode-and-forward networks with the best relay selection scheme. The work in \cite{8493528} derived the coverage probability for a non-orthogonal multiple access-enabled wireless network, where the beam misalignment at both the BS and users was considered. Based on the statistics of the Student's t-distribution, the downlink SNR and rate coverage probability for mmWave cellular networks were developed in \cite{9723625}. The work in \cite{9792581} studied the system performance for a terahertz communication network that utilizes user-centric base station clustering and non-coherent joint transmission with the inclusion of beam misalignment. 
In the field of UAV communications, the authors in \cite{9470921} evaluated the performance of millimeter-wave backhauling in a large-scale aerial-terrestrial network, considering the impact of beam misalignment on the backhaul link. However, the performance analysis of all these works was based on the 2D beam pattern and one dimensional beam misalignment, which can be unsuitable for a 3D terrestrial-aerial wireless network. Hence, the impact of the beam misalignment in a 3D UAV communication network still needs to be addressed\footnote{The work in \cite{linbinbin} investigated network performance with the beam misalignment in a large-scale single-tier UAV communication network, but the considered system model (e.g., the user association scheme and the distribution of beam misalignment) is simple and special.}.

\subsection{Contributions}
In this work, as inspired by the necessities and benefits brought by the multi-tier UAV communications mentioned before, we study the system-level downlink performance for a multi-tier UAV communication network, where the impact of the imperfect beam alignment for 3D beam patterns is included. We further investigate the effects of the number of antennas, the deployment altitude, and the density of UAVs on the outage performance. We make the following contributions in this paper.
\begin{itemize}
\item 
Based on stochastic geometry, a mathematical framework is developed to calculate the outage probability for a multi-tier UAV communication system, where imperfect beam alignment is considered. Note that our developed framework works for two widely adopted user association schemes, called the closest distance association scheme (CDAS) and the maximum average power association scheme (MAPAS). It also covers a special case considered in the literature, i.e., the outage probability for perfect beam alignment.

\item We come up with an analytical tractable random beam misalignment model for our considered 3D beam pattern in this 3D terrestrial-aerial wireless communication network. Based on this, the distribution of the beamforming gain under imperfect beam alignment is characterized, which allows for the effective and feasible computation of the Laplace transform of the probability density function (PDF) of the aggregate interference (i.e., a key factor in calculating the outage probability). 

\item The results show that the outage probability under imperfect beam alignment can be about 100 times larger than the outage probability under perfect beam alignment, which suggests the importance of the stability of the beam in UAV communications. Our results also demonstrate that too many antennas (equivalently, the main-lobe beamwidth is very narrow) are not beneficial to UAV communications under imperfect beam alignment. Moreover, under the case of beam misalignment, the optimum number of antennas needs to be set relatively larger when the UAVs are denser or deployed at higher altitudes for both association schemes.
\end{itemize}
\subsection{Notations and Paper Organization}
This paper adopts the following notation. $\mathcal{\Pr(\cdot)}$ is used to denote the probability measure, $\bar{\cdot}$ is the set complement operator, $\|\cdot\|$ denotes the Euclidean norm of its argument. $\mathbb{E}_{X}[\cdot]$ denotes the expectation operation with regards to the random variable $X$. $f_{X}(x)$ denotes the PDF of $X$. $F_{X}(x)$ denotes the cumulative distribution function (CDF) of $X$. $\mathcal{L}_{X}(s)$ denotes the Laplace transform of the PDF of $X$. The main mathematical symbols and random variables adopted in this work are summarized in TABLE~\ref{tab:var}.

\ifCLASSOPTIONonecolumn
\begin{table}[htbp]
\centering
\caption{Summary of the Main Mathematical Symbols}
\label{tab:var}
\begin{tabular}{|c|c||c|c|}\hline
Symbol &Meaning  & Symbol& Meaning\\
\hline
$K$ &Number of UAV tiers  &$x_{k,i}$ &The $i$-th UAV on the $k$-th tier \\
\hline
$\lambda_{k}$  & UAV density of the $k$-th tier  & $\Phi_{k}$ &The point process of UAVs on the $k$-th tier \\ \hline
${H_{k}}$ &UAV height for the $k$-th tier&$T$& SINR threshold \\ \hline
$\eta$& Type of link ($\eta=\LL$ for LoS; $\eta=\NN$ for NLoS)& $\kappa$ & Node specifier ($\kappa=u$ for UE; $\kappa=v$ for UAV) \\
\hline
  \multirow{2}{*}{\makecell[c]{$\tau$}} & \multirow{2}{*}{\makecell[c]{Type of plane ($\tau=a$ for azimuth plane; \\$\tau=e$ for elevation plane)}} &\multirow{2}{*}{\makecell[c]{$\theta^{(\tau)}_{\kappa}$}} &\multirow{2}{*}{\makecell[c]{Half-power beamwidth for the node $\kappa$ \\
 in the type $\tau$ plane}} \\ 
  &&&\\ \hline
   $\delta_{\kappa}^{(\tau)}$  & Beamsteering error for the node $\kappa$ in the type $\tau$  & $N_{\kappa}$ & Number of antennas for node $\kappa$\\ \hline
 $P_{t,k}$  & Transmit power of the $k$-th tier UAVs &$A_{\eta}$& The additional attenuation factor for the $\eta$-type link \\ \hline
   $G_{j}$ & Beamforming gain of the $j$-th beam alignment case & $G_{\kappa}$, $g_{\kappa}$ &The main-lobe gain and side-lobe gain for the node $\kappa$\\ \hline
  \multirow{2}{*}{\makecell[c]{$h_{\eta,k,i}$}} & \multirow{2}{*}{\makecell[c]{The fading power gain between the UE and \\the $i$-th UAV on the $k$-th tier with $\eta$-type link}} & \multirow{2}{*}{\makecell[c]{$M_{k,j,i}$}} & \multirow{2}{*}{\makecell[c]{The $j$-th type overall beamforming gain on the interfering\\link between the UE and the $i$-th UAV on the $k$-th tier}}\\  &&&\\
  \hline
  $m_{\eta}$ & Nakagami-$m$ parameter for the $\eta$-type link  &$\alpha_{\eta}$ & Path loss coefficient of $\eta$-type link   \\ \hline
 \end{tabular}
\end{table}
\else
\begin{table*}[!t]
\centering
\caption{Summary of the Main Mathematical Symbols}
\label{tab:var}
\begin{tabular}{|c|c||c|c|}\hline
Symbol &Meaning  & Symbol& Meaning\\
\hline
$K$ &Number of UAV tiers  &$x_{k,i}$ &The $i$-th UAV on the $k$-th tier \\
\hline
$\lambda_{k}$  & UAV density of the $k$-th tier  & $\Phi_{k}$ &The point process of UAVs on the $k$-th tier \\ \hline
${H_{k}}$ &UAV height for the $k$-th tier&$T$& SINR threshold \\ \hline
$\eta$& Type of link ($\eta=\LL$ for LoS; $\eta=\NN$ for NLoS)& $\kappa$ & Node specifier ($\kappa=u$ for UE; $\kappa=v$ for UAV) \\
\hline
  \multirow{2}{*}{\makecell[c]{$\tau$}} & \multirow{2}{*}{\makecell[c]{Type of plane ($\tau=a$ for azimuth plane; \\$\tau=e$ for elevation plane)}} &\multirow{2}{*}{\makecell[c]{$\theta^{(\tau)}_{\kappa}$}} &\multirow{2}{*}{\makecell[c]{Half-power beamwidth for the node $\kappa$ \\
 in the type $\tau$ plane}} \\ 
  &&&\\ \hline
   $\delta_{\kappa}^{(\tau)}$  & Beamsteering error for the node $\kappa$ in the type $\tau$ plane & $N_{\kappa}$ & Number of antennas for node $\kappa$\\ \hline
 $P_{t,k}$  & Transmit power of the $k$-th tier UAVs &$A_{\eta}$& The additional attenuation factor for the $\eta$-type link \\ \hline
   $G_{j}$ & Beamforming gain of the $j$-th beam alignment case & $G_{\kappa}$, $g_{\kappa}$ &The main-lobe gain and side-lobe gain for the node $\kappa$\\ \hline
  \multirow{2}{*}{\makecell[c]{$h_{\eta,k,i}$}} & \multirow{2}{*}{\makecell[c]{The fading power gain between the UE and \\the $i$-th UAV on the $k$-th tier with $\eta$-type link}} & \multirow{2}{*}{\makecell[c]{$M_{k,j,i}$}} & \multirow{2}{*}{\makecell[c]{The $j$-th type overall beamforming gain on the interfering\\link between the UE and the $i$-th UAV on the $k$-th tier}}\\  &&&\\
  \hline
  $m_{\eta}$ & Nakagami-$m$ parameter for the $\eta$-type link  &$\alpha_{\eta}$ & Path loss coefficient of $\eta$-type link   \\ \hline
 \end{tabular}
\end{table*}
 \fi

The remainder of this paper is organized as follows. Section~\ref{sec:model} presents the system model and the investigated performance metric of the proposed multi-tier UAV network. The beamforming gain distribution in the presence of random beam misalignment is analyzed in Section~\ref{sec:beam}. Section~\ref{sec:op} characterizes the serving distance and the aggregate interference, and the summary of the outage probability is also presented. In Section~\ref{sec:s&r}, the numerical results are presented to analyze the influence of different parameters on the outage performance. Finally, the paper is concluded in Section~\ref{sec:concl}.

\section{System Model}\label{sec:model}
\subsection{Spatial Model}
We consider a multi-tier UAV communication network~\cite{Hossain-2018,9530405}, where the UAVs act as aerial BSs serving the ground user equipments (UEs). The UAVs are assumed to be distributed on $K$ tiers and the height of each tier is denoted by $H_{k}$, where $k\in{1,2,...,K}$ and $H_1\leq H_2\leq ...\leq H_K$. For the $k$-th ($k=1,2,...,K$) tier, the locations of the UAVs are modeled as an independent homogeneous Poisson point process (HPPP) with density $\lambda_{k}$ on height $H_k$, denoted as $\Phi_k$. Throughout the paper, we refer to $x_{k,i}$ as both the random location and the $i$-th UAV on the $k$-th tier itself. The locations of UEs are also modeled as an independent HPPP on the ground. 
The density of UEs is assumed to be much larger than the density of UAVs \cite{8443416,8761739,9248588}, ensuring that each UAV is connected to at least one UE. Without loss of generality, we focus on the performance at a typical UE, which is assumed to be located at the origin \cite{8856258,9665243,8345639,9723625,9792581,8443416,8761739}. The signal-to-interference-plus-noise ratio (SINR) performance for this typical UE also holds for other UEs according to Slivnyak's theorem~\cite{Haenggi-2012}. The considered system model is depicted in Fig.~\ref{fig:distribution}, where the desired signal link and the interference link are represented by the solid line and dashed line, respectively.

\ifCLASSOPTIONonecolumn
\begin{figure}[h]
\centering
        \includegraphics[width=0.9\linewidth]{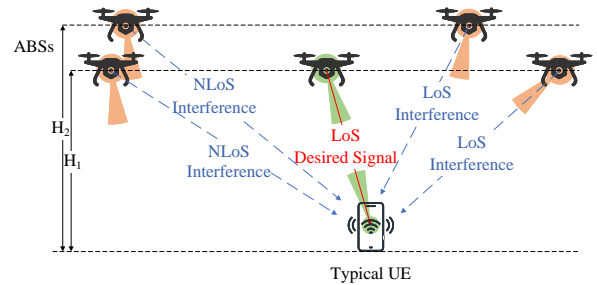}
        \caption{Illustration of the system model.}
        \label{fig:distribution}
\end{figure}
\else
\begin{figure}[h]
\centering
        \includegraphics[width=\linewidth]{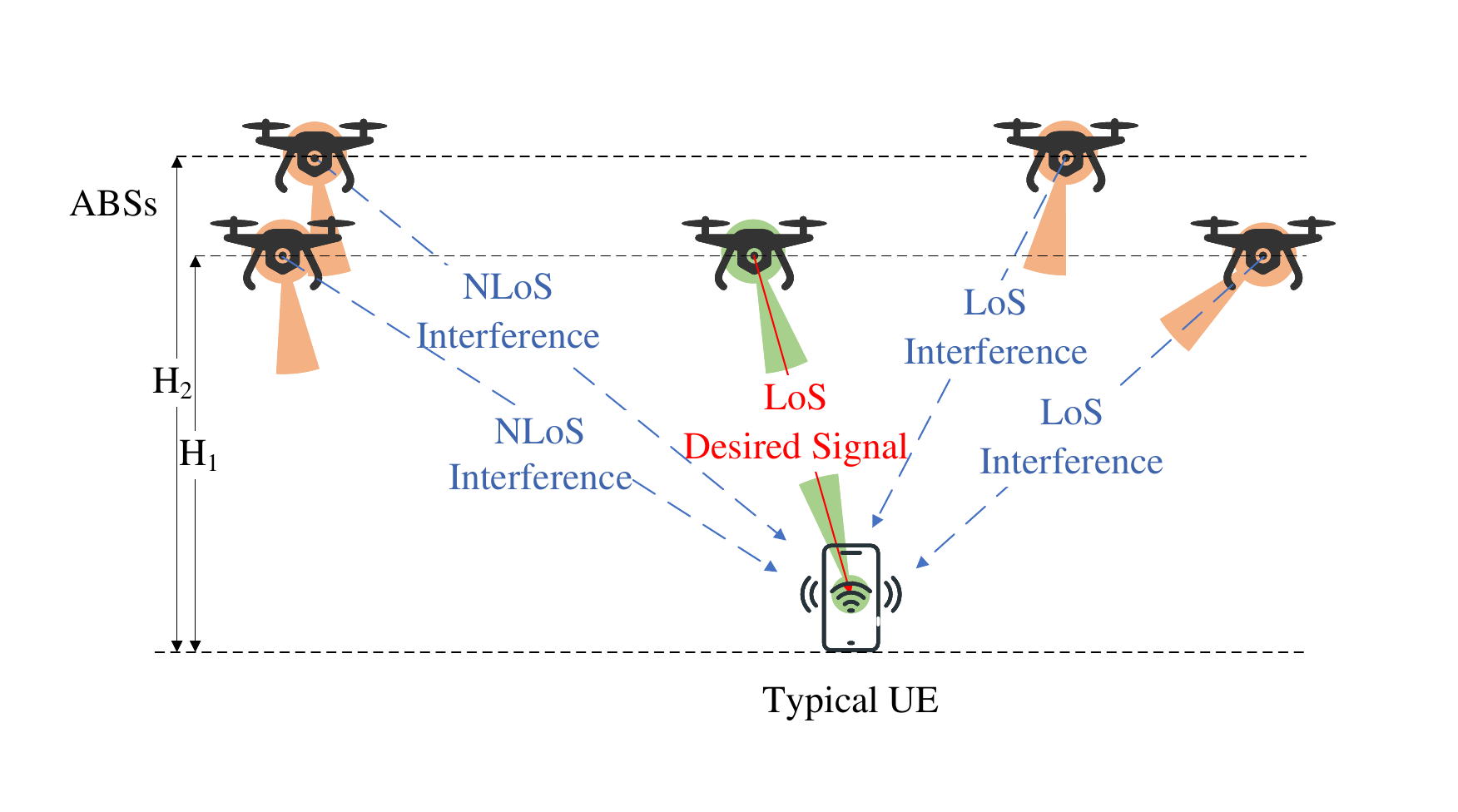}
        \caption{Illustration of the system model.}
        \label{fig:distribution}
\end{figure}
\fi
\subsection{Beamforming and Beam Misalignment Model}\label{subsec:bf model}
\begin{figure}[!t]
\centering
        \includegraphics[width=150pt]{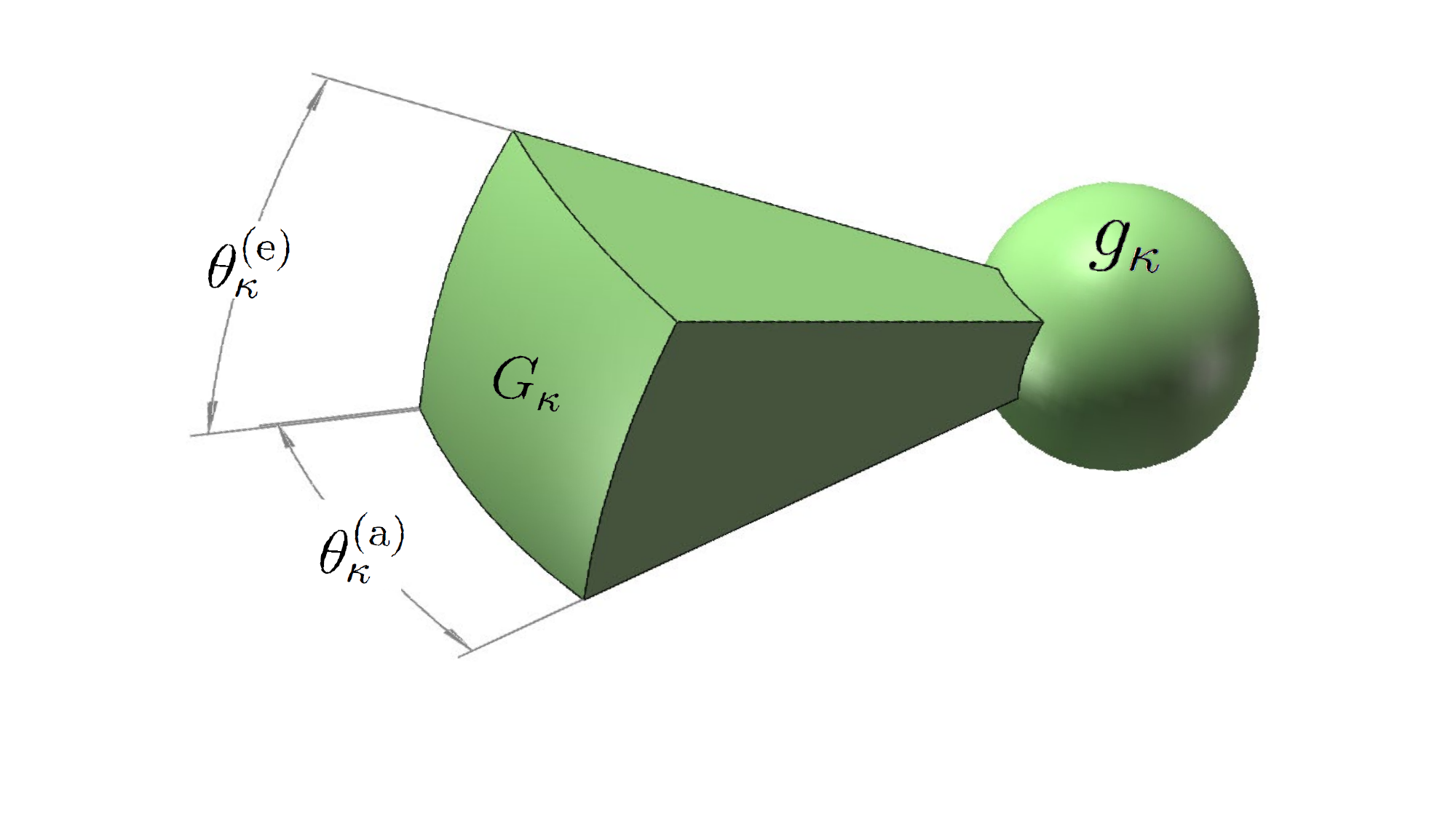}
        \caption{The sectorized antenna pattern.}
        \label{fig:bfmodel}
\end{figure}

To reduce the interference whereby improving the coverage performance, multiple antennas are employed by both UAVs and UEs, enabling the formation of directional beamforming. For analytical tractability, the antenna array pattern for both UAVs and UEs is modeled by a 3D sectorized antenna pattern\footnote{
The purpose of the adoption of the sectorized array pattern is to reduce the computation complexity as shown in the following sections. Note that this sectorized antenna pattern is widely adopted in the literature, e.g., \cite{8856258,9113318,venugopal2016device,9906433}. The consideration of a more complicated antenna array pattern is outside the scope of this work.} as depicted in Fig. \ref{fig:bfmodel}. As shown in the figure, four parameters describe this model, i.e., the main-lobe gain $G_{\kappa}$, side-lobe gain $g_{\kappa}$, the half-power beamwidth in the azimuth $\theta^{(\mathrm{a})}_{\kappa}$ and the half-power beamwidth in the elevation $\theta^{(\mathrm{e})}_{\kappa}$. The subscript $\kappa$ is used to denote the antenna pattern parameters for the UE ($\kappa=u$) and UAV ($\kappa=v$).
Generally, the perfect beam alignment between the UE and its target UAV is highly beneficial to the wireless communication network, which has been widely assumed in the literature. In reality, due to the factors such as airflow, beam misalignment may occur between the UE and UAV (e.g., the main-lobe of the UAV fails to point towards the main-lobe of the UE). Hence, we include the effect of beam misalignment in the system.

The impact of beam misalignment is captured by the beamsteering errors. Since the considered network is a 3D spatial model and the beamforming model is also 3D, we decompose the beamsteering errors into two directions, i.e., the azimuth and the elevation. Fig.~\ref{fig:misalignment} plots the effect of misalignment between the UAV and the typical UE in the azimuth plane and the elevation plane, respectively. For example, in Fig.~\ref{fig:azimuth_misalignment}, the dashed line denotes the direction of error-free boresight (i.e., the ideal alignment) and the solid arrow is the actual boresight. The deviation angle of the actual boresight from the error-free boresight seen from the azimuth plane is denoted by $\delta_{\kappa}^{(a)}$, named as the beamsteering error in azimuth for UAV ($\kappa=v$) or UE ($\kappa=u$). Similarly, as depicted in Fig.~\ref{fig:elevation_misalignment}, the beamsteering error in elevation is $\delta_{\kappa}^{(e)}$. As the factors causing misalignment are unpredictable and random, for analytical tractability, we regard the beamsteering errors as independent random variables. The PDF of beamsteering errors is denoted by $f_{\Delta_{\kappa}^{(\tau)}}\left(\delta_{\kappa}^{(\tau)}\right)$, where $\delta_{\kappa}^{(\tau)}\in[\epsilon_{\min,\kappa}^{(\tau)},\epsilon_{\max,\kappa}^{(\tau)}]$ and the superscript $\tau$ denotes the type of plane (i.e., $\tau=a$ for azimuth plane and $\tau=e$ for elevation plane). Note that $\delta_{\kappa}^{(\tau)}$ takes the positive value if the deviation of the actual boresight from the error-free boresight is clock-wise. Otherwise, it takes a negative value.
\ifCLASSOPTIONonecolumn
\begin{figure}[t]
\centering
\subfigure[Misalignment in the azimuth plane.]{\label{fig:azimuth_misalignment}\includegraphics[width=0.3\textwidth]{fig/azimuth_misalignment_new.pdf}}
\hspace{.5in}
\subfigure[Misalignment in the elevation plane.]{\label{fig:elevation_misalignment}\includegraphics[width=0.3\textwidth]{fig/elevation_misalignment_new.pdf}}
\caption{Illustration of the beam misalignment.}\label{fig:misalignment}
\end{figure}
\else
\begin{figure*}[t]
\centering
\subfigure[Misalignment in the azimuth.]{\label{fig:azimuth_misalignment}\includegraphics[width=125pt]{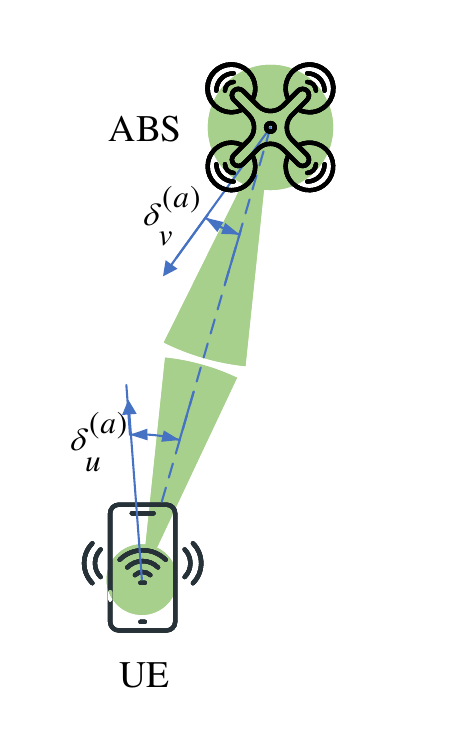}}
\subfigure[Misalignment in the elevation.]{\label{fig:elevation_misalignment}\includegraphics[width=125pt]{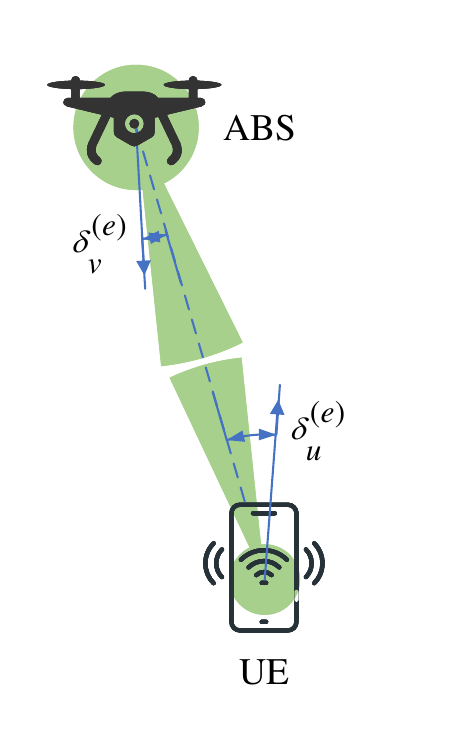}}
\caption{Illustration of the beam misalignment.}\label{fig:misalignment}
\end{figure*}
\fi
\subsection{Channel Model}\label{subsec:channel model}
\ifCLASSOPTIONonecolumn
A power-law path-loss model with small scale block fading is adopted to characterize the transmission channel. In addition, the link between the UAV and the ground UE can be LoS or non-line-of-sight (NLoS). According to \cite{al2014optimal}, the probability of LoS transmission and NLoS transmission are respectively given by $p_{k,\LL}(z) = \frac{1}{{1 + a\exp \left( { - b\left( {\frac{{180}}{\pi }{{\tan }^{ - 1}}\left( {\frac{H_{k}}{z}} \right) - a} \right)} \right)}}$ and $ p_{k,\NN}(z) = 1-p_{k,\LL}(z)$, where $z$ represents the horizontal distance between the UE and the projection point of the UAV on the ground, $a$ and $b$ are parameters related to the transmission environment. Consequently, the received power $P_{\mathrm{r}}(y)$ at the typical UE from the UAV $x_{k,i}$ is
\begin{align}\label{eq:channel}
P_{r}(x_{k,i})= \left\{ \begin{array}{ll}
       P_{t,k} A_{\mathrm{L}} M_{v,k,i} M_{u,k,i} h_{\mathrm{L},k,i}\|x_{k,i}\|^{-\alpha_{\mathrm{L}}}, &{\textrm{LoS with probability } p_{k,\LL}(\sqrt{\|x_{k,i}\|^2-H_k^2}) ;}\\
       P_{t,k} A_{\NN} M_{v,k,i} M_{u,k,i} h_{\NN,k,i}\|x_{k,i}\|^{-\alpha_{\NN}},                &{\textrm{NLoS with probability } p_{k,\NN}(\sqrt{\|x_{k,i}\|^2-H_k^2});}
                    \end{array} \right.
\end{align}
\else
A power-law path-loss model with small scale block fading is adopted to characterize the transmission channel. In addition, the link between the UAV and the ground UE can be LoS or non-line-of-sight (NLoS). According to \cite{al2014optimal}, the probability of LoS transmission and NLoS transmission are respectively given by $p_{k,\LL}(z) = \frac{1}{{1 + a\exp \left( { - b\left( {\frac{{180}}{\pi }{{\tan }^{ - 1}}\left( {\frac{H_{k}}{z}} \right) - a} \right)} \right)}}$ and $ p_{k,\NN}(z) = 1-p_{k,\LL}(z)$, where $z$ represents the horizontal distance between the UE and the projection point of the UAV on the ground, $a$ and $b$ are parameters related to the transmission environment. Consequently, the received power $P_{\mathrm{r}}(x_{k,i})$ at the typical UE from the UAV $x_{k,i}$ is shown in Eq. \eqref{eq:channel} at the top of the next page,
\begin{figure*}[!t]
\begin{align}\label{eq:channel}
P_{r}(x_{k,i})= \left\{ \begin{array}{ll}
       P_{t,k} A_{\mathrm{L}} M_{v,k,i} M_{u,k,i} h_{\mathrm{L},k,i}\|x_{k,i}\|^{-\alpha_{\mathrm{L}}}, &{\textrm{LoS with probability } p_{k,\LL}(\sqrt{\|x_{k,i}\|^2-H_k^2}) ;}\\
       P_{t,k} A_{\NN} M_{v,k,i} M_{u,k,i} h_{\NN,k,i}\|x_{k,i}\|^{-\alpha_{\NN}},                &{\textrm{NLoS with probability } p_{k,\NN}(\sqrt{\|x_{k,i}\|^2-H_k^2}).}
                    \end{array} \right.
\end{align}
\end{figure*}
\fi
where $P_{t,k}$ is the transmit power for UAVs in the $k$-th tier, $\|x_{k,i}\|$ is the distance from $x_{k,i}$ to the typical UE, $M_{v,k,i}$ and $ M_{u,k,i}$ are the beamforming gain from the $x_{k,i}$ and the UE, respectively. Let $\eta$ denote the type of transmission link, i.e., $\eta=\textrm{L}$ for LoS transmission and $\eta=\textrm{N}$ for NLoS transmission. Then, $A_{\eta}$, $\alpha_{\eta}$ and $h_{\eta,k,i}$ are the additional attenuation factor, path-loss exponent and the fading power gain for the $\eta$-type link, respectively. Here, we assume the link suffers from identically and independently distributed fading. Hence, for analytical tractability, similar to \cite{9250029,8422376,8856258,8859647,8833522}, the LoS fading is modeled as the Nakagami-$m$ distribution. Note that the Nakagami-$m$ distribution can approximate the Rician fading \cite{Guo-2017, 9250029}. When $m_N=1$, the Nakagami-$m$ distribution is the Rayleigh fading corresponding to the NLoS link. Therefore, $h_{\eta,k,i}$ follows Gamma distribution with shape parameter $m_{\eta}$ and scale parameter $\frac{1}{m_{\eta}}$ \cite{Guo-2017}, i.e., the PDF of $h_{\eta,k,i}$ is $f_{H_{\eta,k,i}}(h_{\eta,k,i})=\frac{{h_{\eta,k,i}}^{(m_{\eta}-1)} {m_{\eta}}^{m_{\eta}} e^{(-m_{\eta} h_{\eta,k,i})}}{\Gamma(m_{\eta})}$.

\subsection{Performance Metric}
\ifCLASSOPTIONonecolumn
This work adopts the outage probability as the metric to evaluate the UAV communication network performance. The outage probability is defined as the probability that the SINR at the typical UE is less than a certain threshold $T$. Moreover, two widely adopted association schemes are considered in this work. One is the MAPAS (i.e., the user is associated with the UAV providing the maximum average received power), and another one is the CDAS (i.e., the user is associated with the closest UAV). We then can express the outage probability as
\begin{align}
{P_{{\rm{o}}}}=& \Pr\left(\textsf{SINR} < T\right)\notag\\
=& 1  - \sum_{\tilde{\eta}\in\{\LL,\NN\}}\sum_{\tilde{k}=1}^K \Pr\left(\frac{ P_{t,\tilde{k}} A_{\tilde{\eta}} M_{v,\tilde{k},0} M_{u,\tilde{k},0} h_{\tilde{\eta},\tilde{k},0}\|x_{\tilde{k},0}\|^{-\alpha_{\tilde{\eta}}}}{I + \sigma ^2} > T \cap \{ {\tilde{k}\&\tilde{\eta}}\} \right)\nonumber\\
= &1 - \sum_{\tilde{\eta}\in\{\LL,\NN\}}\sum_{\tilde{k} =1}^K\mathbb{E}_{\mathbf{\delta}_{\tilde{k},0},r_{\tilde{k},\tilde{\eta}}}\left[\mathbb{E}_{M_{v,\tilde{k},0},M_{u,\tilde{k},0}}\left[\left.\sum_{t=0}^{m_{\tilde{\eta}}-1} \frac{(-s)^{t}}{t !} \frac{\mathrm{d}^{t}\exp(-s\sigma^2)\mathcal{L}_{I_{{\tilde{k}},\tilde{\eta}}}(s)}{\mathrm{~d} s^{t}}\right|_{s=\frac{Tr_{\tilde{k},\tilde{\eta}}^{ {\alpha _{\tilde{\eta}}}}}{P_{t,\tilde{k}}A_{\tilde{\eta}}M_{v,\tilde{k},0},M_{u,\tilde{k},0}}} \right]\right]     ,
\label{eq:outage_df}
\end{align}
\else
This work adopts the outage probability as the metric to evaluate the performance of a multi-tier UAV communication network. The outage probability is defined as the probability that the SINR at the typical UE is less than a certain threshold $T$. Moreover, two widely adopted association schemes are considered in this work. One is the MAPAS (i.e., the user is associated with the UAV providing the maximum average received power), and another one is the CDAS (i.e., the user is associated with the closest UAV). We then can express the outage probability as Eq. \eqref{eq:outage_df} at the top of the next page,
\begin{figure*}[!t]
\begin{align}
{P_{{\rm{o}}}}=& \Pr\left(\textsf{SINR} < T\right)\notag\\
=& 1  - \sum_{\tilde{\eta}\in\{\LL,\NN\}}\sum_{\tilde{k}=1}^K \Pr\left(\frac{ P_{t,\tilde{k}} A_{\tilde{\eta}} M_{v,\tilde{k},0} M_{u,\tilde{k},0} h_{\tilde{\eta},\tilde{k},0}\|x_{\tilde{k},0}\|^{-\alpha_{\tilde{\eta}}}}{I + \sigma ^2} > T \cap \{ {\tilde{k}\&\tilde{\eta}}\} \right)\nonumber\\
= &1 - \sum_{\tilde{\eta}\in\{\LL,\NN\}}\sum_{\tilde{k} =1}^K\mathbb{E}_{\mathbf{\delta}_{\tilde{k},0},r_{\tilde{k},\tilde{\eta}}}\left[\mathbb{E}_{M_{v,\tilde{k},0},M_{u,\tilde{k},0}}\left[\left.\sum_{t=0}^{m_{\tilde{\eta}}-1} \frac{(-s)^{t}}{t !} \frac{\mathrm{d}^{t}\exp(-s\sigma^2)\mathcal{L}_{I_{{\tilde{k}},\tilde{\eta}}}(s)}{\mathrm{~d} s^{t}}\right|_{s=\frac{Tr_{\tilde{k},\tilde{\eta}}^{ {\alpha _{\tilde{\eta}}}}}{P_{t,\tilde{k}}A_{\tilde{\eta}}M_{v,\tilde{k},0},M_{u,\tilde{k},0}}} \right]\right]   .
\label{eq:outage_df}
\end{align}
\hrulefill
\end{figure*}
\fi
%
where the index $0$ is used to denote the associated UAV and $r_{\tilde{k},\tilde{\eta}}=||x_{\tilde{k},0}||$ is the serving distance between the typical UE and its associated UAV which lies at the $\tilde{k}$-th tier with $\tilde{\eta}$-type link. $I$ is the aggregate interference from all interfering UAVs\footnote{In this work, we focus on investigating the impact of beam misalignment of a UAV communication system; hence, only the interference from UAVs is included. Note that our analysis can be easily extended to include the interference from terrestrial BSs.}. $\mathbf{\delta}_{\tilde{k},0}\triangleq \{\delta_{u}^{(a)},\delta_{u}^{(e)},\delta_{v,\tilde{k},0}^{(a)},\delta_{v,\tilde{k},0}^{(e)}\}$ is the beamsteering error set. $\mathcal{L}_{I_{{\tilde{k}},\tilde{\eta}}}(s)$ is the conditional Laplace transform of the PDF of aggregate interference from UAVs, given that the UE is associated with a $\tilde{k}$-th tier UAV with $\tilde{\eta}$-type link, the serving distance is $r_{\tilde{k},\tilde{\eta}}$ and beamsteering error of the typical UE in the elevation plane is $\delta_{u}^{(e)}$. The last step comes from the fact that $h_{\tilde{\eta},\tilde{k},0}$ follows Gamma distribution with integer shape parameter $m_{\tilde{\eta}}$ (i.e., the parameter in Nagakami-m fading $m_{\tilde{\eta}}$ is assumed to be integer) and $\frac{\textup{d}^t}{\textup{d} s^t}\exp(-s\sigma^2)\mathcal{L}_{I_{{\tilde{k}},\tilde{\eta}}}(s)=\mathbb{E}_{{I_{{\tilde{k}},\tilde{\eta}}}}\left\{\left(-\left(I_{{\tilde{k}},\tilde{\eta}}+\sigma^2\right)\right)^t\exp\left(-s\left(I_{{\tilde{k}},\tilde{\eta}}+\sigma^2\right) \right)\right\}$~\cite{Guo-2017}. Moreover, the beamforming gain and the conditional Laplace transform of the PDF of aggregate interference can be expressed depending on the beamsteering error and the serving distance for analytical tractability; hence, the formula is displayed in two folds of expectation.

 From~Eq. \eqref{eq:outage_df}, the distributions of the serving distance $r_{\tilde{k},\tilde{\eta}}$, the beamforming gains, and the conditional Laplace transform of the PDF of aggregate interference are important, which will be detailed in the following sections.

%

%

\section{Characterization of Beamforming Gain Distribution under Imperfect Beam Alignment}\label{sec:beam}
In this section, we analyze the beamforming gain distribution by considering the imperfect beam alignment. We first characterize the beamforming gain for the typical link (i.e., the link between the typical UE and its associated UAV). The beamforming gain distribution for the interference link, which plays a key role in analyzing the interference, is captured later on.

\subsection{Characterization of Beamforming Gain Distribution for the Typical Link}
In fact, regardless of the typical link and the interference link, there are four possible beam alignment cases as described below:
\begin{itemize}
 \item Case 1: The main-lobe of UAV is aligned with the main-lobe of UE. Hence, the overall beamforming gain on the link is $M_{k,1,i}\triangleq  M_{v,k,i} M_{u,k,i}= G_1\triangleq G_{v}G_{u}$.
 \item Case 2: The main-lobe of UAV is aligned with the side-lobe of UE. $M_{k,2,i}=G_2\triangleq G_{v}g_{u}$.
 \item Case 3: The side-lobe of UAV is aligned with the main-lobe of UE. $M_{k,3,i}=G_3\triangleq g_{v}G_{u}$.
 \item Case 4: The side-lobe of UAV is aligned with the side-lobe of UE. $M_{k,4,i}=G_4\triangleq g_{v}g_{u}$.
\end{itemize}
\ifCLASSOPTIONonecolumn
Under the perfect beam alignment, the overall beamforming gain on the typical link is always equal to $G_1$ (i.e., $G_{v}G_{u}$). However, under the beam misalignment case, the overall beamforming gain on the typical link is a random variable depending on the beamsteering errors. To reflect the dependent relationship between the beamforming gain and the beamsteering errors, we display the beamforming gain distribution for the typical link by the following piece-wise function
\begin{align}
M_{\tilde{k},0}= \left\{ \begin{array}{ll}
       G_1, &{\mathbf{\mathcal{A}}_1 \triangleq \bigcap\limits_{\kappa\in\{u,v\},\tau\in\{a,e\}}\left\{\delta_{\kappa}^{(\tau)}\in\underbrace{\left[\max\left\{\epsilon_{\min,\kappa}^{(\tau)},-\frac{\theta_{\kappa}^{(\tau)}}{2}\right\},\min\left\{\epsilon_{\max,\kappa}^{(\tau)},\frac{\theta_{\kappa}^{(\tau)}}{2}\right\} \right]}_{\mathbf{B}_{\kappa}^{(\tau)} }\right\};}\\
        G_2,   &{\mathbf{\mathcal{A}}_2 \triangleq \bigcap\limits_{\tau\in\{a,e\}}\left\{\delta_{v}^{(\tau)}\in \mathbf{B}_{v}^{(\tau)}\right\}\bigcap
        \left\{ \left\{ \left\{\delta_{u}^{(a)}\in \mathbf{B}_{u}^{(a)}\right\}\bigcap \left\{\delta_{u}^{(e)}\in \bar{\mathbf{B}}_{u}^{(e)}\right\}   \right\} \right.} \\
       \quad &\quad\quad\quad{\left.  \bigcup \left\{ \left\{\delta_{u}^{(a)}\in \bar{\mathbf{B}}_{u}^{(a)}\right\}\bigcap \left\{\delta_{u}^{(e)}\in \mathbf{B}_{u}^{(e)}\right\} \right\}
        \bigcup \left\{ \bigcap\limits_{\tau\in\{a,e\}}\left\{\delta_{u}^{(\tau)}\in \bar{\mathbf{B}}_{u}^{(\tau)}\right\}  \right\}                     \right\};} \\
         G_3,   &{\mathbf{\mathcal{A}}_3 \triangleq \bigcap\limits_{\tau\in\{a,e\}}\left\{\delta_{u}^{(\tau)}\in \mathbf{B}_{u}^{(\tau)}\right\}\bigcap
        \left\{ \left\{ \left\{\delta_{v}^{(a)}\in \mathbf{B}_{v}^{(a)}\right\}\bigcap \left\{\delta_{v}^{(e)}\in \bar{\mathbf{B}}_{v}^{(e)}\right\}   \right\} \right.} \\
       \quad &\quad\quad\quad{\left.  \bigcup \left\{ \left\{\delta_{v}^{(a)}\in \bar{\mathbf{B}}_{v}^{(a)}\right\}\bigcap \left\{\delta_{v}^{(e)}\in \mathbf{B}_{v}^{(e)}\right\} \right\}
        \bigcup \left\{ \bigcap\limits_{\tau\in\{a,e\}}\left\{\delta_{v}^{(\tau)}\in \bar{\mathbf{B}}_{v}^{(\tau)}\right\}  \right\}                     \right\};}\\
          G_4,   &{\mathcal{A}_4 \triangleq \bar{\mathbf{\mathcal{A}}}_1\backslash \{\mathcal{A}_2 \bigcup\mathcal{A}_3\}    ;}
                    \end{array} \right.\label{eq:gain_typical}
\end{align}
where $\bar{\cdot}$ is the set complement operator.
\else
Under the perfect beam alignment, the overall beamforming gain on the typical link is always equal to $G_1$ (i.e., $G_{v}G_{u}$). However, under the beam misalignment case, the overall beamforming gain on the typical link is a random variable depending on the beamsteering errors. To reflect the dependent relationship between the beamforming gain and the beamsteering errors, we display the beamforming gain distribution for the typical link by the piece-wise function as shown in~Eq. \eqref{eq:gain_typical} at the top of the next page, where $\bar{\cdot}$ is the set complement operator.
\begin{figure*}[!h]
\normalsize
\begin{small}
\begin{align}
M_{\tilde{k},0}= \left\{ \begin{array}{ll}
       G_1, &{\mathbf{\mathcal{A}}_1 \triangleq \bigcap\limits_{\kappa\in\{u,v\},\tau\in\{a,e\}}\left\{\delta_{\kappa}^{(\tau)}\in\underbrace{\left[\max\left\{\epsilon_{\min,\kappa}^{(\tau)},-\frac{\theta_{\kappa}^{(\tau)}}{2}\right\},\min\left\{\epsilon_{\max,\kappa}^{(\tau)},\frac{\theta_{\kappa}^{(\tau)}}{2}\right\} \right]}_{\mathbf{B}_{\kappa}^{(\tau)} }\right\};}\\
        G_2,   &{\mathbf{\mathcal{A}}_2 \triangleq \bigcap\limits_{\tau\in\{a,e\}}\left\{\delta_{v}^{(\tau)}\in \mathbf{B}_{v}^{(\tau)}\right\}\bigcap
        \left\{ \left\{ \left\{\delta_{u}^{(a)}\in \mathbf{B}_{u}^{(a)}\right\}\bigcap \left\{\delta_{u}^{(e)}\in \bar{\mathbf{B}}_{u}^{(e)}\right\}   \right\} \right.} \\
       \quad &\quad\quad\quad{\left.  \bigcup \left\{ \left\{\delta_{u}^{(a)}\in \bar{\mathbf{B}}_{u}^{(a)}\right\}\bigcap \left\{\delta_{u}^{(e)}\in \mathbf{B}_{u}^{(e)}\right\} \right\}
        \bigcup \left\{ \bigcap\limits_{\tau\in\{a,e\}}\left\{\delta_{u}^{(\tau)}\in \bar{\mathbf{B}}_{u}^{(\tau)}\right\}  \right\}                     \right\};} \\
         G_3,   &{\mathbf{\mathcal{A}}_3 \triangleq \bigcap\limits_{\tau\in\{a,e\}}\left\{\delta_{u}^{(\tau)}\in \mathbf{B}_{u}^{(\tau)}\right\}\bigcap
        \left\{ \left\{ \left\{\delta_{v}^{(a)}\in \mathbf{B}_{v}^{(a)}\right\}\bigcap \left\{\delta_{v}^{(e)}\in \bar{\mathbf{B}}_{v}^{(e)}\right\}   \right\} \right.} \\
       \quad &\quad\quad\quad{\left.  \bigcup \left\{ \left\{\delta_{v}^{(a)}\in \bar{\mathbf{B}}_{v}^{(a)}\right\}\bigcap \left\{\delta_{v}^{(e)}\in \mathbf{B}_{v}^{(e)}\right\} \right\}
        \bigcup \left\{ \bigcap\limits_{\tau\in\{a,e\}}\left\{\delta_{v}^{(\tau)}\in \bar{\mathbf{B}}_{v}^{(\tau)}\right\}  \right\}                     \right\};}\\
          G_4,   &{\mathcal{A}_4 \triangleq \bar{\mathbf{\mathcal{A}}}_1\backslash \{\mathcal{A}_2 \bigcup\mathcal{A}_3\}    .}
                    \end{array} \right.\label{eq:gain_typical}
\end{align}
\end{small}
\hrulefill
\vspace*{4pt}
\end{figure*}
\fi

 The beamforming gain distribution for the typical link is obtained in the following way. When the beamforming gain of the typical link contributed from $\kappa$ is the main-lobe gain $G_{\kappa}$,  it requires the beamsteering errors in both planes must be less than the half-power beamwidth on both planes, i.e., $|\delta_{\kappa}^{(\tau)}|\leq \frac{\theta_{\kappa}^{(\tau)}}{2}$, $\forall \tau, \kappa$. Moreover, $\delta_{\kappa}^{(\tau)}$ is always bounded within $[\epsilon_{\min,\kappa}^{(\tau)},\epsilon_{\max,\kappa}^{(\tau)}]$. Let $\mathbf{B}_{\kappa}^{(\tau)}$ represent the allowable range of misalignment, that is, the misalignment within $\mathbf{B}_{\kappa}^{(\tau)}$ is still regarded as main lobe alignment. Hence, $\forall \kappa$, $\delta_{\kappa}^{(\tau)}\in\mathbf{B}_{\kappa}^{(\tau)}$ implies that the beamforming gain from $\kappa$ is $G_{\kappa}$. Otherwise, the beamforming gain is the side-lobe gain $g_{\kappa}$. By considering all possible combinations, we arrive at the beamforming gain distribution for the typical link in~Eq. \eqref{eq:gain_typical}.

\subsection{Characterization of Beamforming Gain Distribution for the Interference Link}\label{subsec:Infofmisalign}
For analytical tractability, rather than directly expressing the beamforming gain distribution for the interference link as shown above, we categorize the interfering UAVs into different point processes (PPs) depending on the overall beamforming gain on the interfering link, and then characterize the distributions of these categorized PPs.

Since there are four possible beam alignment cases, for any $k$-th tier, the original PP $\Phi_k$ can be decomposed into four Poisson point processes (PPPs), denoted as $\Phi_{k,j}$, $j=1,2,3,4$. The key features of these PPPs are described by TABLE \ref{tab:align}, where ${p_{{\rm{{t,m}}}}} = \frac{{\theta^{(\mathrm{a})}_{v}\theta^{(\mathrm{e})}_{v}}}{{{{\rm{\pi }}^2}}}$, and $\mathcal{S}_k(r_{\tilde{k},\tilde{\eta}},\delta_{u}^{(e)},\theta^{(\mathrm{a})}_{u},\theta^{(\mathrm{e})}_{u})$ denotes the projection region on the UAV tier of height $H_k$ formed by the main-lobe of the UE when the UE is associated to a $\tilde{k}$-th tier UAV with $\tilde{\eta}$-type link and the beamsteering errors of the UE are $\delta_{u}^{(a)}$ and $\delta_{u}^{(e)}$. The formulation of the project region is presented in Lemma 1.
\ifCLASSOPTIONonecolumn
\begin{table}[!t]
\centering
\caption{Key features for the decomposed PPPs of interfering UAVs on the k-th tier}
\label{tab:align}
\begin{tabular}{|c|c|c|c|c|}\hline
Categories $\Phi_{k,j}$ & $\Phi_{k,1}$ & $\Phi_{k,2}$ & $\Phi_{k,3}$ & $\Phi_{k,4}$ \\
\hline
 \multirow{2}{*}{\makecell[c]{ Overall beamforming gain\\on the interfering link $M_{k,j,i}$}}   & \multirow{2}{*}{$G_1\triangleq G_{v}G_{u}$} &\multirow{2}{*}{ $G_2\triangleq G_{v}g_{u}$} & \multirow{2}{*}{$G_3\triangleq g_{v}G_{u}$} & \multirow{2}{*}{$G_4\triangleq g_{v}g_{u}$}\\
 & & & &\\\hline
Density  $\lambda_{k,j} $ & $p_{{\rm{{t,m}}}}\lambda_{k}$ & $p_{{\rm{{t,m}}}}\lambda_{k}$ & $(1-p_{{\rm{{t,m}}}})\lambda_{k}$ & $(1-p_{{\rm{{t,m}}}})\lambda_{k}$\\ \hline
\multirow{3}{*}{Region}   & \multirow{3}{*}{\makecell[c]{ $\mathcal{S}_k(r_{\tilde{k},\tilde{\eta}},\delta_{u}^{(e)},$\\ $\theta^{(\mathrm{a})}_{u},\theta^{(\mathrm{e})}_{u})$ \\on height $H_k$}} & \multirow{3}{*}{\makecell[c]{ $\mathbb{R}^2\backslash\mathcal{S}_k(r_{\tilde{k},\tilde{\eta}},\delta_{u}^{(e)},$\\ $\theta^{(\mathrm{a})}_{u},\theta^{(\mathrm{e})}_{u})$ \\on height $H_k$}} &\multirow{3}{*}{\makecell[c]{ $\mathcal{S}_k(r_{\tilde{k},\tilde{\eta}},\delta_{u}^{(e)},$\\ $\theta^{(\mathrm{a})}_{u},\theta^{(\mathrm{e})}_{u})$ \\on height $H_k$}}  &\multirow{3}{*}{\makecell[c]{ $\mathbb{R}^2\backslash\mathcal{S}_k(r_{\tilde{k},\tilde{\eta}},\delta_{u}^{(e)},$\\ $\theta^{(\mathrm{a})}_{u},\theta^{(\mathrm{e})}_{u})$ \\on height $H_k$}}\\
 &&&&\\
 &&&&\\\hline
\end{tabular}
\end{table}
\else
\begin{table*}[!t]
\centering
\caption{Key features for the decomposed PPPs of interfering UAVs on the k-th tier}
\label{tab:align}
\begin{tabular}{|c|c|c|c|c|}\hline
Categories $\Phi_{k,j}$ & $\Phi_{k,1}$ & $\Phi_{k,2}$ & $\Phi_{k,3}$ & $\Phi_{k,4}$ \\
\hline
 \multirow{2}{*}{\makecell[c]{ Overall beamforming gain\\on the interfering link $M_{k,j,i}$}}   & \multirow{2}{*}{$G_1\triangleq G_{v}G_{u}$} &\multirow{2}{*}{ $G_2\triangleq G_{v}g_{u}$} & \multirow{2}{*}{$G_3\triangleq g_{v}G_{u}$} & \multirow{2}{*}{$G_4\triangleq g_{v}g_{u}$}\\
 & & & &\\\hline
Density  $\lambda_{k,j} $ & $p_{{\rm{{t,m}}}}\lambda_{k}$ & $p_{{\rm{{t,m}}}}\lambda_{k}$ & $(1-p_{{\rm{{t,m}}}})\lambda_{k}$ & $(1-p_{{\rm{{t,m}}}})\lambda_{k}$\\ \hline
&&&&\\
\multirow{3}{*}{Region}   & \multirow{3}{*}{\makecell[c]{ $\mathcal{S}_k(r_{\tilde{k},\tilde{\eta}},\delta_{u}^{(e)},$\\ $\theta^{(\mathrm{a})}_{u},\theta^{(\mathrm{e})}_{u})$ \\on height $H_k$}} & \multirow{3}{*}{\makecell[c]{ $\mathbb{R}^2\backslash\mathcal{S}_k(r_{\tilde{k},\tilde{\eta}},\delta_{u}^{(e)},$\\ $\theta^{(\mathrm{a})}_{u},\theta^{(\mathrm{e})}_{u})$ \\on height $H_k$}} &\multirow{3}{*}{\makecell[c]{ $\mathcal{S}_k(r_{\tilde{k},\tilde{\eta}},\delta_{u}^{(e)},$\\ $\theta^{(\mathrm{a})}_{u},\theta^{(\mathrm{e})}_{u})$ \\on height $H_k$}}  &\multirow{3}{*}{\makecell[c]{ $\mathbb{R}^2\backslash\mathcal{S}_k(r_{\tilde{k},\tilde{\eta}},\delta_{u}^{(e)},$\\ $\theta^{(\mathrm{a})}_{u},\theta^{(\mathrm{e})}_{u})$ \\on height $H_k$}}\\
 &&&&\\
  &&&&\\
 &&&&\\\hline
\end{tabular}
\end{table*}
\fi

The categorization of the interfering UAVs is obtained in the following. We first characterize the beamforming gain contributed by the interfering UAVs. Similar to \cite{khan2016millimeter,9470921,bai2014coverage,8856258}, we assume that the beam orientation in the azimuth direction is uniformly distributed in the range of $[0,2\pi)$ and the beam orientation in the elevation direction is uniformly distributed in the range of $[0, \pi/2)$ for analytical tractability. Note that when the typical UE is located within the half-power beamwidth of the UAV in both azimuth and elevation directions, the beamforming gain on the interfering link is $G_{v}$; otherwise, the beamforming gain is $g_{v}$. Consequently, the probability for the beamforming gain on the interfering link being main-lobe gain $G_v$, denoted as $p_{\rm{t,m}}$, is given by
\begin{align}
{p_{{\rm{{t,m}}}}}{\rm{ = }}\frac{\theta^{(\mathrm{a})}_{v}}{{2{\rm{\pi }}}} \cdot \frac{{{\theta^{(\mathrm{e})}_{v}}}}{{{\rm{\pi }}/2}} = \frac{{\theta^{(\mathrm{a})}_{v}\theta^{(\mathrm{e})}_{v}}}{{{{\rm{\pi }}^2}}},
\label{eq:ml_prob}
\end{align}
and the probability for the beamforming gain on the interfering link being side-lobe gain $g_v$ is $1- p_{\rm{t,m}}$. According to the thinning theorem in stochastic geometry \cite{Haenggi-2012}, the independent thinning applied to a PPP generates a new PPP. Hence, it can be regarded that the PP of the interfering UAVs with $G_{v}$ beamforming gain is an HPPP with density $p_{{\rm{{t,m}}}}\lambda_{k}$ and the PP of the interfering UAVs with $g_{v}$ beamforming gain is an HPPP with density $(1-p_{{\rm{{t,m}}}})\lambda_{k}$.

Based on the above decomposed PPs, we further characterize the beamforming gain contributed from the typical UE. Clearly, when the UAV is located in region $\mathcal{S}_k(r_{\tilde{k},\tilde{\eta}},\delta_{u}^{(e)},\theta^{(\mathrm{a})}_{u},\theta^{(\mathrm{e})}_{u})$ on height $H_k$, the beamforming gain from the typical UE is $G_{u}$. Otherwise, the beamforming gain is $g_{u}$. Hence, the aforementioned PPs can be further divided. For example, for those UAVs whose beamforming gain to the typical UE is $G_{v}$, their locations can be further decomposed into two PPs depending on the residing region, i.e., $\Phi_{k,1}$ and $\Phi_{k,2}$. $\Phi_{k,1}$ is a PPP with density $p_{{\rm{{t,m}}}}\lambda_{k}$ in region $\mathcal{S}_k(r_{\tilde{k},\tilde{\eta}},\delta_{u}^{(e)},\theta^{(\mathrm{a})}_{u},\theta^{(\mathrm{e})}_{u})$ on height $H_k$, which constitutes the UAVs with overall beamforming gain $G_1$ on the interfering link. $\Phi_{k,2}$ is a PPP with density $p_{{\rm{{t,m}}}}\lambda_{k}$ in region $\mathbb{R}^2\backslash\mathcal{S}_k(r_{\tilde{k},\tilde{\eta}},\delta_{u}^{(e)},\theta^{(\mathrm{a})}_{u},\theta^{(\mathrm{e})}_{u})$ on height $H_k$, which constitutes the UAVs with overall beamforming gain $G_2$ on the interfering link. $\Phi_{k,3}$ and $\Phi_{k,4}$ can be found in the same way.

\begin{lemma}
Based on the considered 3D sectorized antenna pattern in Section II-B, given that the UE is associated with the $\tilde{k}$-th tier UAV with $\tilde{\eta}$-type link (i.e., the serving distance is $r_{\tilde{k},\tilde{\eta}}$) and the project area formed by the main-lobe of the UE on the $k$-th tier, $\mathcal{S}_k(r_{\tilde{k},\tilde{\eta}},\delta_{u}^{(e)},\theta^{(\mathrm{a})}_{u},\theta^{(\mathrm{e})}_{u})$, can be approximated by a ring sector, which is specified by an angle $\theta^{(\mathrm{a})}_{u}$, an inner radius $Z_{in,\tilde{k},k}$ and an outer radius $Z_{out,\tilde{k},k}$, as shown in the dashed area of Fig.~\ref{fig:proj_a_new}. The formulations of $Z_{in,\tilde{k},k}$ and $Z_{out,\tilde{k},k}$ are respectively given by  
\ifCLASSOPTIONonecolumn
    \begin{align}
        {Z_{in,\tilde{k},k}} = \left\{ {\begin{array}{*{20}{ll}}
        {\frac{H_k}{{\tan (\beta  + {\delta_{u}^{(e)}} + \frac{{{\theta^{(e)}_{u}}}}{2})}}},&{\beta  + {\delta_{u}^{(e)}} < \frac{{\rm{\pi }}}{2} - \frac{{{\theta^{(e)}_{u}}}}{2}};\\
        0,&{\textup{otherwise}};
        \end{array}}\label{eq:z_in} \right.
    \end{align}

    \begin{align}
        {Z_{out,\tilde{k},k}} = \left\{ {\begin{array}{*{20}{ll}}
        {\frac{H_k}{{\tan (\beta  + {\delta_{u}^{(e)}} - \frac{{{\theta^{(e)}_{u}}}}{2})}}},&{\frac{{{\theta^{(e)}_{u}}}}{2} < \beta  + {\delta_{u}^{(e)}} < \frac{{\rm{\pi }}}{2} - \frac{{{\theta^{(e)}_{u}}}}{2}};\\
        {\frac{H_k}{{\tan (\frac{{\rm{\pi }}}{2} - {\theta^{(e)}_{u}})}}},&{\beta  + {\delta_{u}^{(e)}} \geq \frac{{\rm{\pi }}}{2} - \frac{{{\theta^{(e)}_{u}}}}{2}};\\
        \infty, &{\textup{otherwise}};
        \end{array}}\label{eq:z_out} \right.
    \end{align}
    \else
      \begin{align}
        {Z_{in,\tilde{k},k}} \!\!=\!\! \left\{ {\begin{array}{*{20}{ll}}
        \!\!\!{\frac{H_k}{{\tan (\beta  + {\delta_{u}^{(e)}} + \frac{{{\theta^{(e)}_{u}}}}{2})}}},\!\!\!\!\!&{\beta  + {\delta_{u}^{(e)}} \!\!<\!\! \frac{{\rm{\pi }}}{2} - \frac{{{\theta^{(e)}_{u}}}}{2}};\\
        \!\!\!0,\!\!\!\!\!&{\textup{otherwise}};
        \end{array}}\label{eq:z_in} \right.
    \end{align}

    \begin{align}
        {Z_{out,\tilde{k},k}} \!\!= \!\!\left\{ {\begin{array}{*{20}{ll}}
        \!\!\!{\frac{H_k}{{\tan (\beta  + {\delta_{u}^{(e)}} - \frac{{{\theta^{(e)}_{u}}}}{2})}}},\!\!\!\!\!&{\frac{{{\theta^{(e)}_{u}}}}{2}\!\! <\!\! \beta  + {\delta_{u}^{(e)}} \!\!<\!\! \frac{{\rm{\pi }}}{2} - \frac{{{\theta^{(e)}_{u}}}}{2}};\\
        \!\!\!{\frac{H_k}{{\tan (\frac{{\rm{\pi }}}{2} - {\theta^{(e)}_{u}})}}},\!\!\!\!\!&{\beta  + {\delta_{u}^{(e)}} \!\!\geq\!\! \frac{{\rm{\pi }}}{2} - \frac{{{\theta^{(e)}_{u}}}}{2}};\\
        \!\!\!\infty,\!\!\!\!\! &{\textup{otherwise}};
        \end{array}}\label{eq:z_out} \right.
    \end{align}
    \fi
    where $\beta=\arcsin \left( \frac{H_{\tilde{k}}}{r_{\tilde{k},\tilde{\eta}}} \right)$ represents the elevation angle between the typical UE and the associated UAV.

\ifCLASSOPTIONonecolumn
\begin{figure}[htbp]
    \centering
    \subfigure[Project area $\mathcal{S}_k(r_{\tilde{k},\tilde{\eta}},\delta_{u}^{(e)},\theta^{(\mathrm{a})}_{u},\theta^{(\mathrm{e})}_{u})$ in the azimuth plane.]{\label{fig:proj_a_new}\includegraphics[width=165pt]{fig/proj_a_new_small.pdf}} \ \ \
    \subfigure[Projection in the elevation plane (Take the UAV tier number $K=2$ as an example, and the typical UE is associated with the red UAV).]{\label{fig:proj_z}\includegraphics[width=235pt]{fig/proj_z_new_smaller_again_cunzaipiancha_R1.pdf}}
    \caption{Beam projection of the UE's main-lobe beam.}
        \label{fig:proj}
    \end{figure}
\else
\begin{figure*}[htbp]
    \centering
    \subfigure[Project area $\mathcal{S}_k(r_{\tilde{k},\tilde{\eta}},\delta_{u}^{(e)},\theta^{(\mathrm{a})}_{u},\theta^{(\mathrm{e})}_{u})$ in the azimuth plane.]{\label{fig:proj_a_new}\includegraphics[width=165pt]{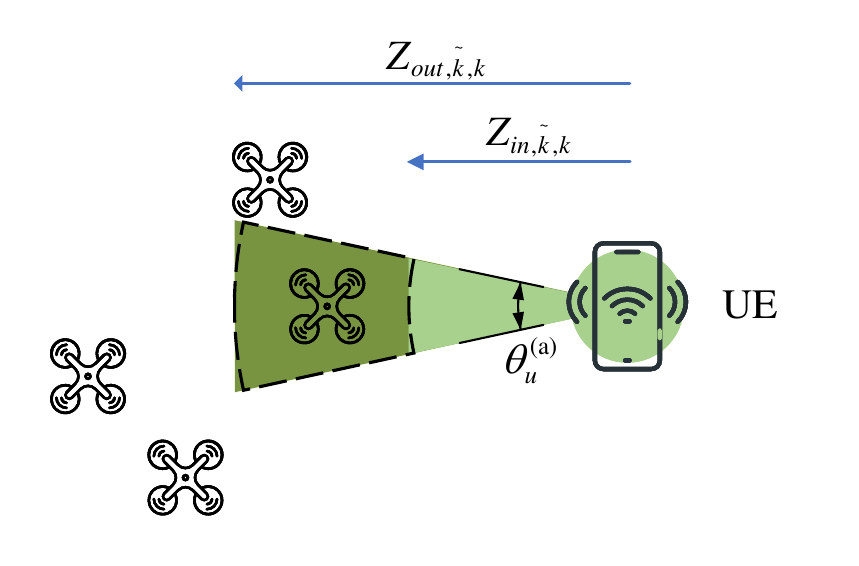}} \ \ \
    \subfigure[Projection in the elevation plane (Take the UAV tier number $K=2$ as an example, and the typical UE is associated with the red UAV).]{\label{fig:proj_z}\includegraphics[width=235pt]{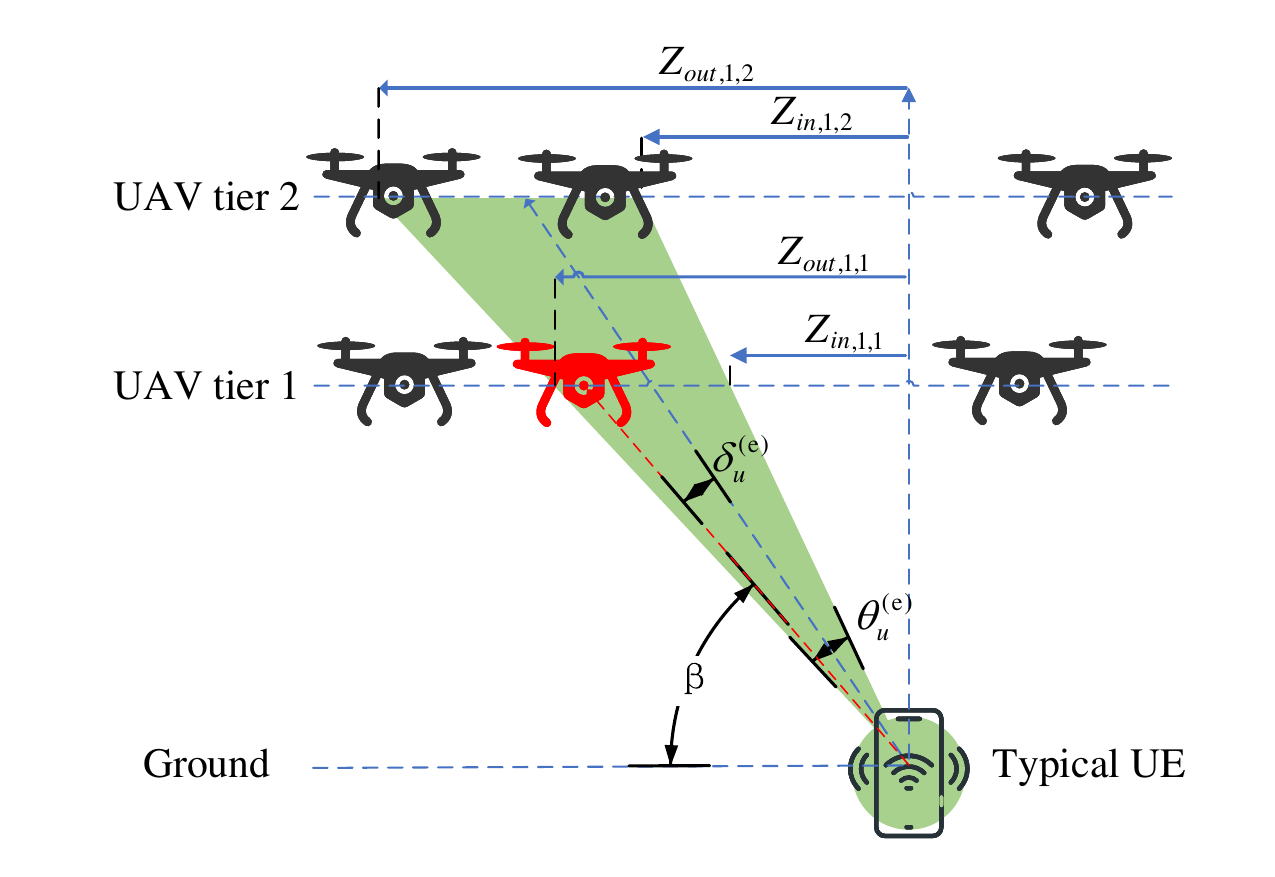}}
    \caption{Beam projection of the UE's main-lobe beam.}
        \label{fig:proj}
    \end{figure*}
\fi
\begin{IEEEproof}
 See Appendix A.
\end{IEEEproof}
\end{lemma}

\section{Characterization of the Serving Distance and the Aggregate Interference}\label{sec:op}
In this section, we first present the analysis for the distribution of the serving distance (i.e., the distance between the typical UE and its associated UAV) and the Laplace transform of the PDF of aggregate interference. Finally, the overall outage probability is summarized.

\subsection{Distribution of the Serving Distance}
For our considered association schemes, i.e., MAPAS and CDAS, according to the stochastic geometry, the distribution of the serving distance are presented in the following Lemmas.
\ifCLASSOPTIONonecolumn
\begin{lemma}
Under the MAPAS, the PDF of serving distance with $\tilde{\eta}$-type link is given by
\begin{align}\label{eq:PDF_R_N_result}
{f}_{R_{\tilde{k},\tilde{\eta}}}(r_{\tilde{k},\tilde{\eta}})= 
     &2{\rm{\pi }}{\lambda_{\tilde{k}}}{r_{\tilde{k},\tilde{\eta}}}p_{{\tilde{k}},\tilde{\eta}}\left(\sqrt {{r_{\tilde{k},\tilde{\eta}}}^2 - {H_{\tilde{k}}}^2} \right)\exp \left( { - 2{\rm{\pi }}{\lambda_{\tilde{k}}}\int_0^{\sqrt {{r_{\tilde{k},\tilde{\eta}}}^2 - {H_{\tilde{k}}}^2} } {z{p_{{\tilde{k}},\tilde{\eta}}}(z){\rm{d}}z} } \right) \notag\\
    &\times\exp \left(-2 \pi \!\!\!\!\!\sum\limits_{k \in\{1,2, \ldots K\}} \!\!\!\!\!\lambda_{k} \int_{0}^{{Z_{eq,k}}(r_{\tilde{k},\tilde{\eta}},\eta_s)} p_{k,{\eta_s}}(z) z \mathrm{d} z -2 \pi\!\!\!\!\! \sum\limits_{k \in\{1,2, \ldots, K\} \setminus \tilde{k}}\!\!\!\!\! \lambda_{k} \int_{0}^{{Z_{eq,k}}(r_{\tilde{k},\tilde{\eta}},\tilde{\eta})}p_{k,\tilde{\eta}}(z) z \mathrm{d} z\right) , 
\end{align}
where $\tilde{\eta}$ represents the type of the association link, $\eta_s=\LL$ if $\tilde{\eta}=\NN$, $\eta_s=\NN$ if $\tilde{\eta}=\LL$, and the equivalent distance  ${Z_{eq,k}}(r_{\tilde{k},\tilde{\eta}},\eta)$ represents the closest horizontal distance between the interfering UAVs on the $k$-th tier with $\eta$-type link and the typical UE when the typical UE is associated with a specific UAV with distance $r_{\tilde{k},\tilde{\eta}}$. It is expressed as
\begin{equation}
\label{eq:closest dis}
{Z_{eq,k}}(r_{\tilde{k},\tilde{\eta}},\eta) =\begin{cases}
    \sqrt {\max \left\{ {0,{{\left( {\frac{{{A_{\rm{L}}}}}{{{A_{\rm{N}}}}}} \right)}^{ - \frac{2}{{{\alpha _{\rm{N}}}}}}}{r_{\tilde{k},\tilde{\eta}}^{\frac{{2{\alpha _{\rm{L}}}}}{{{\alpha _{\rm{N}}}}}}}\!\!\! - \!\!{H_k}^2} \right\}} ,&\tilde{\eta} = {\rm{L\& }}\eta = {\rm{N}};\\
    \sqrt {\max \left\{ {0,{{\left( {\frac{{{A_{\rm{N}}}}}{{{A_{\rm{L}}}}}} \right)}^{ - \frac{2}{{{\alpha _{\rm{L}}}}}}}{r_{\tilde{k},\tilde{\eta}}^{\frac{{2{\alpha _{\rm{N}}}}}{{{\alpha _{\rm{L}}}}}}}\!\!\! -\!\! {H_k}^2} \right\}} ,&\tilde{\eta} = {\rm{N\& }}\eta = {\rm{L}};\\
    \sqrt {\max \left\{ {0,{r_{\tilde{k},\tilde{\eta}}^2} - {H_k}^2} \right\}} ,&\textup{otherwise}.
\end{cases}
\end{equation}
\end{lemma}
\begin{lemma}
Under the CDAS, the PDF of the serving distance is given by
\begin{align}\label{eq:PDF_CDAS}
{f}_{R_{\tilde{k},\tilde{\eta}}}(r_{\tilde{k},\tilde{\eta}})=
     2{\rm{\pi }}{\lambda_{\tilde{k}}}{r_{\tilde{k},\tilde{\eta}}} p_{\tilde{k},\tilde{\eta}}\left(\sqrt {{r_{\tilde{k},\tilde{\eta}}}^2 - {H_{\tilde{k}}}^2} \right)\exp \left(- \pi \!\!\!\!\!\!\sum_{{k} \in\{1,2, \ldots K\}}\!\!\!\!\! \lambda_{k}{Z^2_{eq,k}}(r_{\tilde{k},\tilde{\eta}})\right),
\end{align}
where the equivalent distance ${Z_{eq,k}}(r_{\tilde{k},\tilde{\eta}})$ is $\sqrt{\max \left\{ {0,{r_{\tilde{k},\tilde{\eta}}^2} - {H_k}^2} \right\}}$.
\end{lemma}
\begin{IEEEproof}
See Appendix B.
\end{IEEEproof}
\else
\begin{lemma}
Under the MAPAS, the PDF of serving distance with $\tilde{\eta}$-type link is given by Eq. \eqref{eq:PDF_R_N_result} at the top of the next page,
\begin{figure*}[!t]
\normalsize
\begin{align}\label{eq:PDF_R_N_result}
{f}_{R_{\tilde{k},\tilde{\eta}}}(r_{\tilde{k},\tilde{\eta}})= 
     &2{\rm{\pi }}{\lambda_{\tilde{k}}}{r_{\tilde{k},\tilde{\eta}}}p_{{\tilde{k}},\tilde{\eta}}\left(\sqrt {{r_{\tilde{k},\tilde{\eta}}}^2 - {H_{\tilde{k}}}^2} \right)\exp \left( { - 2{\rm{\pi }}{\lambda_{\tilde{k}}}\int_0^{\sqrt {{r_{\tilde{k},\tilde{\eta}}}^2 - {H_{\tilde{k}}}^2} } {z{p_{{\tilde{k}},\tilde{\eta}}}(z){\rm{d}}z} } \right) \notag\\
    &\times\exp \left(-2 \pi \!\!\!\!\!\sum\limits_{k \in\{1,2, \ldots K\}} \!\!\!\!\!\lambda_{k} \int_{0}^{{Z_{eq,k}}(r_{\tilde{k},\tilde{\eta}},\eta_s)}  \!\!\!\!\!\!\!\!\!\! p_{k,{\eta_s}}(z) z \mathrm{d} z -2 \pi\!\!\!\!\! \sum\limits_{k \in\{1,2, \ldots, K\} \setminus \tilde{k}}\!\!\!\!\! \lambda_{k} \int_{0}^{{Z_{eq,k}}(r_{\tilde{k},\tilde{\eta}},\tilde{\eta})} \!\!\!\!\! \!\!\!\!\!p_{k,\tilde{\eta}}(z) z \mathrm{d} z\right) . 
\end{align}
\end{figure*}
where $\tilde{\eta}$ represents the type of the association link, $\eta_s=\LL$ if $\tilde{\eta}=\NN$, $\eta_s=\NN$ if $\tilde{\eta}=\LL$, and the equivalent distance  ${Z_{eq,k}}(r_{\tilde{k},\tilde{\eta}},\eta)$ represents the closest horizontal distance between the interfering UAVs on the $k$-th tier with $\eta$-type link and the typical UE, when the typical UE is associated with a specific UAV with distance $r_{\tilde{k},\tilde{\eta}}$.
It is expressed as
\begin{align}
\label{eq:closest dis}
&{Z_{eq,k}}(r_{\tilde{k},\tilde{\eta}},\eta)=\notag\\ 
&\begin{cases}
    \!\!\!\sqrt {\max \left\{ {0,{{\left( {\frac{{{A_{\rm{L}}}}}{{{A_{\rm{N}}}}}} \right)}^{ - \frac{2}{{{\alpha _{\rm{N}}}}}}}{r_{\tilde{k},\tilde{\eta}}^{\frac{{2{\alpha _{\rm{L}}}}}{{{\alpha _{\rm{N}}}}}}}\!\!\! - \!\!{H_k}^2} \right\}} ,\!\!\!\!\!\!&\tilde{\eta} = {\rm{L\& }}\eta = {\rm{N}};\\
    \!\!\!\sqrt {\max \left\{ {0,{{\left( {\frac{{{A_{\rm{N}}}}}{{{A_{\rm{L}}}}}} \right)}^{ - \frac{2}{{{\alpha _{\rm{L}}}}}}}{r_{\tilde{k},\tilde{\eta}}^{\frac{{2{\alpha _{\rm{N}}}}}{{{\alpha _{\rm{L}}}}}}}\!\!\! -\!\! {H_k}^2} \right\}} ,\!\!\!\!\!\!&\tilde{\eta} = {\rm{N\& }}\eta = {\rm{L}};\\
    \!\!\!\sqrt {\max \left\{ {0,{r_{\tilde{k},\tilde{\eta}}^2} - {H_k}^2} \right\}} ,\!\!\!\!\!\!&\textup{otherwise}.
\end{cases}
\end{align}
\end{lemma}
\begin{lemma}
Under the CDAS, the PDF of serving distance is given by Eq. \eqref{eq:PDF_CDAS} at the top of this page,
\begin{figure*}[!t]
\normalsize
\begin{align}\label{eq:PDF_CDAS}
{f}_{R_{\tilde{k},\tilde{\eta}}}(r_{\tilde{k},\tilde{\eta}})=
     2{\rm{\pi }}{\lambda_{\tilde{k}}}{r_{\tilde{k},\tilde{\eta}}} p_{\tilde{k},\tilde{\eta}}\left(\sqrt {{r_{\tilde{k},\tilde{\eta}}}^2 - {H_{\tilde{k}}}^2} \right)\exp \left(- \pi \!\!\!\!\!\!\sum_{{k} \in\{1,2, \ldots K\}}\!\!\!\!\! \lambda_{k}{Z^2_{eq,k}}(r_{\tilde{k},\tilde{\eta}})\right).
\end{align}
\end{figure*}
where the equivalent distance ${Z_{eq,k}}(r_{\tilde{k},\tilde{\eta}})$ is $\sqrt{\max \left\{ {0,{r_{\tilde{k},\tilde{\eta}}^2} - {H_k}^2} \right\}}$.
\end{lemma}
\begin{IEEEproof}
See Appendix B.
\end{IEEEproof}
\fi
\subsection{Laplace Transform of the PDF of Aggregate Interference}
According to the decomposed PPs mentioned in Section III-B, we can further express the aggregate interference as
\begin{align}
{I_{{\tilde{k}},\tilde{\eta}}}&\!\!=\!\!\sum\limits_{k=1}^{K}\!\!\sum\limits_{j=1}^{4}\!\!\sum\limits_{\eta \in {\LL,\NN}}\sum\limits_{x_{k,i} \in \Phi_{k,j,\eta}\setminus x_{\tilde{k},0}} \!\!\!\!\!\!\!\!\!\!\!\! A_{\eta} P_{\mathrm{t},k} M_{k,j,i} h_{\eta,j,i} \|x_{k,i}\|^{-\alpha_{\eta}}\!\!,
\end{align}
where ${\Phi _{k,j,{\LL}}}$ denotes the set of LoS link UAVs in ${\Phi_{k,j}}$ and ${\Phi _{k,j,{\NN}}}$ denotes the set of NLoS link UAVs in ${\Phi_{k,j}}$.

Following the definitions of the Laplace transform and the probability generating functional (PGFL) in stochastic geometry, we have the Laplace transform of the PDF of aggregate interference given in the following proposition.
\begin{proposition}\label{prop:L_I}
\ifCLASSOPTIONonecolumn
Given that the typical UE is associated with a $\tilde{k}$-th tier UAV with $\tilde{\eta}$-type link, the serving distance is $r_{\tilde{k},\tilde{\eta}}$ and the beamsteering error of the typical UE in the elevation plane is $\delta_{u}^{(e)}$, the conditional Laplace transform of the PDF of aggregate interference can be expressed as
\begin{align}
\label{eq:L_I}
    \begin{array}{l}
    \mathcal{L}_{I_{{\tilde{k}},\tilde{\eta}}}(s)=
 \prod\limits_{k=1}^{K}\prod\limits_{j=1}^{4}\prod\limits_{\eta \in {\LL,\NN}} \mathbb{E}_{{\Phi_{k,j,\eta}},h_{\eta,j,i}}\left[\prod\limits_{{x_{k,i}} \in {\Phi_{k,j,\eta}\setminus x_{\tilde{k},0}}} \exp \left(-s A_{\eta} P_{\mathrm{t},k} M_{k,j,i} h_{\eta,j,i} {\|x_{k,i}\|}^{-\alpha_{\eta}}\right)\right],
    \end{array}
\end{align}
where the sub-term can be represented by
\begin{align}\label{eq:signal}
&\mathbb{E}_{{\Phi_{k,j,\eta}},h_{\eta,j,i}}\left[ {\prod\limits_{x_{k,i} \in {\Phi _{k,j,\eta}\setminus{x_{\tilde{k},0}}}} {\exp ( - s{A_{\eta}}{P_{t,k}}{M_{k,j,i}}{h_{\eta,j,i}}{\|x_{k,i}\|}^{ - {\alpha _{\eta}}})} } \right] \notag\\&= \left\{ \begin{array}{ll}
       \exp\left[ { - \theta^{(\mathrm{a})}_{u}\lambda_{k,j} \int_{\max \left\{{Z_{in,\tilde{k},k}},{{Z_{eq,k}}(r_{\tilde{k},\tilde{\eta}},\eta)}\right\}}^{\max \left\{{Z_{out,\tilde{k},k}},{{Z_{eq,k}}(r_{\tilde{k},\tilde{\eta}},\eta)}\right\}} \left(1 - \mathcal{M}\left(\eta,k,j,i,z\right)\right){p_{k,\eta}}(z)z{\rm{d}}z} \right], &j=1,3;\\
       \exp\left[  - \left(2{\rm{\pi }}-\theta^{(\mathrm{a})}_{u}\right)\lambda_{k,j}\int_{{{Z_{eq,k}}(r_{\tilde{k},\tilde{\eta}},\eta)}}^\infty  \left(1 - \mathcal{M}\left(\eta,k,j,i,z\right) \right){p_{k,\eta}}(z)z{\rm{d}}z \right.\\
    \left. - \theta^{(\mathrm{a})}_{u}\lambda_{k,j}\int_{\min \left\{{Z_{in,\tilde{k},k}},{{Z_{eq,k}}(r_{\tilde{k},\tilde{\eta}},\eta)}\right\}}^{{Z_{in,\tilde{k},k}}} \left(1 - \mathcal{M}\left(\eta,k,j,i,z\right)\right){p_{k,\eta}}(z)z{\rm{d}}z \right.\\ 
    \left. - \theta^{(\mathrm{a})}_{u}\lambda_{k,j}\int_{\max \left\{{Z_{out,\tilde{k},k}},{{Z_{eq,k}}(r_{\tilde{k},\tilde{\eta}},\eta)}\right\}}^\infty  \left(1 - \mathcal{M}\left(\eta,k,j,i,z\right)\right){p_{k,\eta}}(z)z{\rm{d}}z \right],                &j=2,4;
                    \end{array} \right.
\end{align}
\else
Given that the typical UE is associated with a $\tilde{k}$-th tier UAV with $\tilde{\eta}$-type link, the serving distance is $r_{\tilde{k},\tilde{\eta}}$ and the beamsteering error of the typical UE in the elevation plane is $\delta_{u}^{(e)}$, the conditional Laplace transform of the PDF of aggregate interference can be expressed as Eq. \eqref{eq:L_I} at the top of this page,
\begin{figure*}[!t]
\normalsize
\begin{align}
\label{eq:L_I}
    \begin{array}{l}
    \mathcal{L}_{I_{{\tilde{k}},\tilde{\eta}}}(s)=
 \prod\limits_{k=1}^{K}\prod\limits_{j=1}^{4}\prod\limits_{\eta \in {\LL,\NN}} \mathbb{E}_{{\Phi_{k,j,\eta}},h_{\eta,j,i}}\left[\prod\limits_{{x_{k,i}} \in {\Phi_{k,j,\eta}\setminus x_{\tilde{k},0}}} \exp \left(-s A_{\eta} P_{\mathrm{t},k} M_{k,j,i} h_{\eta,j,i} {\|x_{k,i}\|}^{-\alpha_{\eta}}\right)\right].
    \end{array}
\end{align}
\hrulefill
\end{figure*}
where the sub-term can be represented by Eq. \eqref{eq:signal} at the top of the next page,
\begin{figure*}[!t]
\normalsize
\begin{align}\label{eq:signal}
&\mathbb{E}_{{\Phi_{k,j,\eta}},h_{\eta,j,i}}\left[ {\prod\limits_{x_{k,i} \in {\Phi _{k,j,\eta}\setminus{x_{\tilde{k},0}}}} {\exp ( - s{A_{\eta}}{P_{t,k}}{M_{k,j,i}}{h_{\eta,j,i}}{\|x_{k,i}\|}^{ - {\alpha _{\eta}}})} } \right] \notag\\&= \left\{ \begin{array}{ll}
       \exp\left[ { - \theta^{(\mathrm{a})}_{u}\lambda_{k,j} \int_{\max \left\{{Z_{in,\tilde{k},k}},{{Z_{eq,k}}(r_{\tilde{k},\tilde{\eta}},\eta)}\right\}}^{\max \left\{{Z_{out,\tilde{k},k}},{{Z_{eq,k}}(r_{\tilde{k},\tilde{\eta}},\eta)}\right\}} \left(1 - \mathcal{M}\left(\eta,k,j,i,z\right)\right){p_{k,\eta}}(z)z{\rm{d}}z} \right], &j=1,3;\\
       \exp\left[  - \left(2{\rm{\pi }}-\theta^{(\mathrm{a})}_{u}\right)\lambda_{k,j}\int_{{{Z_{eq,k}}(r_{\tilde{k},\tilde{\eta}},\eta)}}^\infty  \left(1 - \mathcal{M}\left(\eta,k,j,i,z\right) \right){p_{k,\eta}}(z)z{\rm{d}}z \right.\\
    \left. - \theta^{(\mathrm{a})}_{u}\lambda_{k,j}\int_{\min \left\{{Z_{in,\tilde{k},k}},{{Z_{eq,k}}(r_{\tilde{k},\tilde{\eta}},\eta)}\right\}}^{{Z_{in,\tilde{k},k}}} \left(1 - \mathcal{M}\left(\eta,k,j,i,z\right)\right){p_{k,\eta}}(z)z{\rm{d}}z \right.\\ 
    \left. - \theta^{(\mathrm{a})}_{u}\lambda_{k,j}\int_{\max \left\{{Z_{out,\tilde{k},k}},{{Z_{eq,k}}(r_{\tilde{k},\tilde{\eta}},\eta)}\right\}}^\infty  \left(1 - \mathcal{M}\left(\eta,k,j,i,z\right)\right){p_{k,\eta}}(z)z{\rm{d}}z \right],                &j=2,4.
                    \end{array} \right.
\end{align}
\end{figure*}
\fi
where $\mathcal{M}\left(\eta,k,j,i,z\right)\triangleq\left(\frac{{{m_{\eta}}}}{{{m_{\eta}} + s{A_{\eta}}{P_{t,k}}{M_{k,j,i}}{\left(\sqrt{z^2+H_{k}^2}\right)}^{ - \alpha_{\eta} }}} \right)^{{m_{\eta}}}$.
\end{proposition}
\begin{IEEEproof}
See Appendix C.
\end{IEEEproof}
\begin{remark}\label{rm:conditional cp}
With the derived conditional Laplace transform of the PDF of aggregate interference, following the formulation of Eq. \eqref{eq:outage_df}, we can obtain the conditional coverage probability which is conditioned on $\tilde{\eta}$, $\tilde{k}$, $r_{\tilde{k},\tilde{\eta}}$ and $\mathbf{\delta}_{\tilde{k},0}$ as
\ifCLASSOPTIONonecolumn
\begin{align}
\label{eq:P_o}
\begin{array}{l}
\mathcal{\Pr}_{\rm{cov}|\tilde{k},\tilde{\eta}}(s_{\tilde{k},\tilde{\eta},\tilde{j}})=\sum\limits_{t=0}^{m_{\tilde{\eta}}-1} \frac{(-s_{\tilde{k},\tilde{\eta},\tilde{j}})^{t}}{t !} \frac{\mathrm{d}^{t}\exp\left(-s_{\tilde{k},\tilde{\eta},\tilde{j}}\sigma^2\right)\mathcal{L}_{I,\tilde{k},\tilde{\eta}}(s_{\tilde{k},\tilde{\eta},\tilde{j}})}{\mathrm{~d} s_{\tilde{k},\tilde{\eta},\tilde{j}}^{t}}, 
\end{array}
\end{align}
\else
\begin{align}
\label{eq:P_o}
&\mathcal{\Pr}_{\rm{cov}|\tilde{k},\tilde{\eta}}(s_{\tilde{k},\tilde{\eta},\tilde{j}})\notag \\
&=\sum\limits_{t=0}^{m_{\tilde{\eta}}-1} \frac{(-s_{\tilde{k},\tilde{\eta},\tilde{j}})^{t}}{t !} \frac{\mathrm{d}^{t}\exp\left(-s_{\tilde{k},\tilde{\eta},\tilde{j}}\sigma^2\right)\mathcal{L}_{I,\tilde{k},\tilde{\eta}}(s_{\tilde{k},\tilde{\eta},\tilde{j}})}{\mathrm{~d} s_{\tilde{k},\tilde{\eta},\tilde{j}}^{t}}, 
\end{align}
\fi
where $s_{\tilde{k},\tilde{\eta},\tilde{j}}=\frac{Tr_{\tilde{k},\tilde{\eta}}^{ {\alpha _{\tilde{\eta}}}}}{P_{t,\tilde{k}}A_{\tilde{\eta}}G_{\tilde{j}}}$ and $\tilde{j}$ represents the beam alignment case on the association link.
\end{remark}
\subsection{Summary of the Overall Outage Probability}\label{subsec:overallop}

According to Eq. \eqref{eq:outage_df} and the above derivations, we can summarize the overall outage probability as
\ifCLASSOPTIONonecolumn
\begin{align}
{P_{{\rm{o}}}}= 1\! -\!\!\!\!\! \sum_{\tilde{\eta}\in\{\LL,\NN\}}\sum_{\tilde{k} =1}^K \int_{H_{\tilde{k}}}^{\infty} \mathcal{\Pr}\left(\textsf{SINR} > T|\tilde{\eta},\tilde{k},r_{\tilde{k},\tilde{\eta}}\right) {f}_{R_{\tilde{k},\tilde{\eta}}}(r_{\tilde{k},\tilde{\eta}}) \mathrm{d} r_{\tilde{k},\tilde{\eta}}   ,
\label{eq:outage_overall}
\end{align}
\else
\begin{align}
{P_{{\rm{o}}}}= &1\! -\!\!\!\!\! \sum_{\tilde{\eta}\in\{\LL,\NN\}}\sum_{\tilde{k} =1}^K \int_{H_{\tilde{k}}}^{\infty} \mathcal{\Pr}\left(\textsf{SINR} > T|\tilde{\eta},\tilde{k},r_{\tilde{k},\tilde{\eta}}\right)\notag\\ &\times{f}_{R_{\tilde{k},\tilde{\eta}}}(r_{\tilde{k},\tilde{\eta}}) \mathrm{d} r_{\tilde{k},\tilde{\eta}}   ,
\label{eq:outage_overall}
\end{align}
\fi
\ifCLASSOPTIONonecolumn
where $\mathcal{\Pr}\left(\textsf{SINR} > T|\tilde{\eta},\tilde{k},r_{\tilde{k},\tilde{\eta}}\right)$ is the conditional coverage probability given that UE is associated with a $\tilde{k}$-th tier UAV with $\tilde{\eta}$-type link and the serving distance is $r_{\tilde{k},\tilde{\eta}}$. ${f}_{R_{\tilde{k},\tilde{\eta}}}(r_{\tilde{k},\tilde{\eta}})$ is the PDF of serving distance denoted by Eq. \eqref{eq:PDF_R_N_result} for MAPAS or Eq. \eqref{eq:PDF_CDAS} for CDAS. This conditional coverage probability is related to the conditional coverage probability $\mathcal{\Pr}_{\rm{cov}|\tilde{k},\tilde{\eta}}(s_{\tilde{k},\tilde{\eta},\tilde{j}})$ mentioned in Remark~\ref{rm:conditional cp} by averaging $\mathcal{\Pr}_{\rm{cov}|\tilde{k},\tilde{\eta}}(s_{\tilde{k},\tilde{\eta},\tilde{j}})$ over the distribution of beamsteering errors (i.e., removing the condition of beamsteering error set $\mathbf{\delta}_{\tilde{k},0}$). Then, we have $\mathcal{\Pr}\left(\textsf{SINR} >T|\tilde{\eta},\tilde{k},r_{\tilde{k},\tilde{\eta}}\right)$ written as 
\begin{align}
\label{eq:P_dis}
   & \mathcal{\Pr}\left(\textsf{SINR} > T|\tilde{\eta},\tilde{k},r_{\tilde{k},\tilde{\eta}}\right) \notag
    \\
    =&\sum\limits_{\tilde{j}=1}^{4}\int_{{\mathbf{\mathcal{A}}_{\tilde{j}}}}
    \mathcal{\Pr}_{\rm{cov}|\tilde{k},\tilde{\eta}}(s_{\tilde{k},\tilde{\eta},\tilde{j}})f_{\Delta_{u}^{(a)}}(\delta_{u}^{(a)})f_{\Delta_{u}^{(e)}}(\delta_{u}^{(e)})f_{\Delta_{v}^{(a)}}(\delta_{v}^{(a)})f_{\Delta_{v}^{(e)}}(\delta_{v}^{(e)})\rm{d}\mathbf{\delta}_{\tilde{k},0} \notag
    \\
    =&\Omega_{\delta_{v}^{(a)}}\Omega_{\delta_{v}^{(e)}}\Omega_{\delta_{u}^{(a)}}\!\!\!\int_{\max\{\epsilon_{\min,u}^{(e)},-\frac{\theta^{(e)}_{u}}{2}\}}^{\min\{{\epsilon_{\max,u}^{(e)}},\frac{\theta^{(e)}_{u}}{2}\}} \!\!\!\!\!\mathcal{\Pr}_{\rm{cov}|\tilde{k},\tilde{\eta}}(s_{\tilde{k},\tilde{\eta},1})f_{\Delta_{u}^{(e)}}(\delta_{u}^{(e)}) \mathrm{d} \delta_{u}^{(e)}
    \!\!\!+\!\!\Omega_{\delta_{v}^{(a)}}\Omega_{\delta_{v}^{(e)}}\!\!\!\left( \int_{\epsilon_{\min,u}^{(e)}}^{\max\{\epsilon_{\min,u}^{(e)},-\frac{\theta^{(e)}_{u}}{2}\}}\!\!\!\!\!\!\!\!\!\!\!\!\!\!\!\!\!\!\!\!\!\!\!\!\!\!\!\!\!\!\!\!\!\!\! \mathcal{\Pr}_{\rm{cov}|\tilde{k},\tilde{\eta}}(s_{\tilde{k},\tilde{\eta},2})f_{\Delta_{u}^{(e)}}(\delta_{u}^{(e)}) \mathrm{d} \delta_{u}^{(e)} \right.\notag \\&\left.
    +\!\!\!\int_{\min\{{\epsilon_{\max,u}^{(e)}},\frac{\theta^{(e)}_{u}}{2}\}}^{\epsilon_{\max,u}^{(e)}}\!\!\!\! \mathcal{\Pr}_{\rm{cov}|\tilde{k},\tilde{\eta}}(s_{\tilde{k},\tilde{\eta},2})f_{\Delta_{u}^{(e)}}(\delta_{u}^{(e)}) \mathrm{d} \delta_{u}^{(e)}
    \!\!\!+\!\!\left(\!1\!\!-\Omega_{\delta_{u}^{(a)}}\!\right)\!\!\!\!\int_{\max\{\epsilon_{\min,u}^{(e)},-\frac{\theta^{(e)}_{u}}{2}\}}^{\min\{{\epsilon_{\max,u}^{(e)}},\frac{\theta^{(e)}_{u}}{2}\}}\!\! \mathcal{\Pr}_{\rm{cov}|\tilde{k},\tilde{\eta}}(s_{\tilde{k},\tilde{\eta},2})f_{\Delta_{u}^{(e)}}(\delta_{u}^{(e)}) \mathrm{d} \delta_{u}^{(e)} \right)\notag
    \\
    &+\left(1-\Omega_{\delta_{v}^{(a)}}\Omega_{\delta_{v}^{(e)}}\right)\Omega_{\delta_{u}^{(a)}}\int_{\max\{\epsilon_{\min,u}^{(e)},-\frac{\theta^{(e)}_{u}}{2}\}}^{\min\{{\epsilon_{\max,u}^{(e)}},\frac{\theta^{(e)}_{u}}{2}\}} \mathcal{\Pr}_{\rm{cov}|\tilde{k},\tilde{\eta}}(s_{\tilde{k},\tilde{\eta},3})f_{\Delta_{u}^{(e)}}(\delta_{u}^{(e)}) \mathrm{d} \delta_{u}^{(e)}
    \notag \\&
    +\left(1-\Omega_{\delta_{v}^{(a)}}\Omega_{\delta_{v}^{(e)}}\right)\left( \int_{\epsilon_{\min,u}^{(e)}}^{\max\{\epsilon_{\min,u}^{(e)},-\frac{\theta^{(e)}_{u}}{2}\}} \mathcal{\Pr}_{\rm{cov}|\tilde{k},\tilde{\eta}}(s_{\tilde{k},\tilde{\eta},4})f_{\Delta_{u}^{(e)}}(\delta_{u}^{(e)}) \mathrm{d} \delta_{u}^{(e)}\right.\notag 
    \\
    &\left.+\!\!\int_{\min\{{\epsilon_{\max,u}^{(e)}},\frac{\theta^{(e)}_{u}}{2}\}}^{\epsilon_{\max,u}^{(e)}}\!\!\!\! \mathcal{\Pr}_{\rm{cov}|\tilde{k},\tilde{\eta}}(s_{\tilde{k},\tilde{\eta},4})f_{\Delta_{u}^{(e)}}(\delta_{u}^{(e)}) \mathrm{d} \delta_{u}^{(e)}
    \!\!+\!\!\left(\!1\!\!-\!\Omega_{\delta_{u}^{(a)}}\!\right)\!\!\int_{\max\{\epsilon_{\min,u}^{(e)},-\frac{\theta^{(e)}_{u}}{2}\}}^{\min\{{\epsilon_{\max,u}^{(e)}},\frac{\theta^{(e)}_{u}}{2}\}}\!\!\! \mathcal{\Pr}_{\rm{cov}|\tilde{k},\tilde{\eta}}(s_{\tilde{k},\tilde{\eta},4})f_{\Delta_{u}^{(e)}}(\delta_{u}^{(e)}) \mathrm{d} \delta_{u}^{(e)} \!\!\right),
\end{align}
where $s_{\tilde{k},\tilde{\eta},\tilde{j}}=\frac{Tr_{\tilde{k},\tilde{\eta}}^{ {\alpha _{\tilde{\eta}}}}}{P_{t,\tilde{k}}A_{\tilde{\eta}}G_{\tilde{j}}}$,  $\Omega_{\delta_{\kappa}^{(\tau)}} \triangleq F_{\Delta_{\kappa}^{(\tau)}}(\theta^{(\mathrm{\tau})}_{\kappa}/2)-F_{{\Delta_{\kappa}^{(\tau)}}}(- \theta^{(\mathrm{\tau})}_{\kappa}/2)$ and $F_{\Delta_{\kappa}^{(\tau)}}(\delta_{\kappa}^{(\tau)})$ represents the CDF of $\delta_{\kappa}^{(\tau)}$.
\else
where $\mathcal{\Pr}\left(\textsf{SINR} > T|\tilde{\eta},\tilde{k},r_{\tilde{k},\tilde{\eta}}\right)$ is the conditional coverage probability given that UE is associated with a $\tilde{k}$-th tier UAV with $\tilde{\eta}$-type link and the serving distance is $r_{\tilde{k},\tilde{\eta}}$. ${f}_{R_{\tilde{k},\tilde{\eta}}}(r_{\tilde{k},\tilde{\eta}})$ is the PDF of serving distance denoted by Eq. \eqref{eq:PDF_R_N_result} for MAPAS or Eq. \eqref{eq:PDF_CDAS} for CDAS. This conditional coverage probability is related to the conditional coverage probability $\mathcal{\Pr}_{\rm{cov}|\tilde{k},\tilde{\eta}}(s_{\tilde{k},\tilde{\eta},\tilde{j}})$ mentioned in Remark~\ref{rm:conditional cp} by averaging $\mathcal{\Pr}_{\rm{cov}|\tilde{k},\tilde{\eta}}(s_{\tilde{k},\tilde{\eta},\tilde{j}})$ over the distribution of beamsteering errors (i.e., removing the condition of beamsteering error set $\mathbf{\delta}_{\tilde{k},0}$). Then, we have $\mathcal{\Pr}\left(\textsf{SINR} >T|\tilde{\eta},\tilde{k},r_{\tilde{k},\tilde{\eta}}\right)$ written as Eq. \eqref{eq:P_dis} at the top of the next page,
\begin{figure*}[!t]
\centering
\begin{align}
\label{eq:P_dis}
   & \mathcal{\Pr}\left(\textsf{SINR} > T|\tilde{\eta},\tilde{k},r_{\tilde{k},\tilde{\eta}}\right) \notag
    \\
    =&\sum\limits_{\tilde{j}=1}^{4}\int_{{\mathbf{\mathcal{A}}_{\tilde{j}}}}
    \mathcal{\Pr}_{\rm{cov}|\tilde{k},\tilde{\eta}}(s_{\tilde{k},\tilde{\eta},\tilde{j}})f_{\Delta_{u}^{(a)}}(\delta_{u}^{(a)})f_{\Delta_{u}^{(e)}}(\delta_{u}^{(e)})f_{\Delta_{v}^{(a)}}(\delta_{v}^{(a)})f_{\Delta_{v}^{(e)}}(\delta_{v}^{(e)})\rm{d}\mathbf{\delta}_{\tilde{k},0} \notag
    \\
    =&\Omega_{\delta_{v}^{(a)}}\Omega_{\delta_{v}^{(e)}}\Omega_{\delta_{u}^{(a)}}\int_{\max\{\epsilon_{\min,u}^{(e)},-\frac{\theta^{(e)}_{u}}{2}\}}^{\min\{{\epsilon_{\max,u}^{(e)}},\frac{\theta^{(e)}_{u}}{2}\}} \mathcal{\Pr}_{\rm{cov}|\tilde{k},\tilde{\eta}}(s_{\tilde{k},\tilde{\eta},1})f_{\Delta_{u}^{(e)}}(\delta_{u}^{(e)}) \mathrm{d} \delta_{u}^{(e)}
    +\Omega_{\delta_{v}^{(a)}}\Omega_{\delta_{v}^{(e)}}\left( \int_{\epsilon_{\min,u}^{(e)}}^{\max\{\epsilon_{\min,u}^{(e)},-\frac{\theta^{(e)}_{u}}{2}\}}\!\!\!\!\!\!\!\!\!\!\!\!\!\!\!\!\!\!\!\!\!\!\!\!\!\!\!\!\!\!\!\!\!\!\! \mathcal{\Pr}_{\rm{cov}|\tilde{k},\tilde{\eta}}(s_{\tilde{k},\tilde{\eta},2})f_{\Delta_{u}^{(e)}}(\delta_{u}^{(e)}) \mathrm{d} \delta_{u}^{(e)} \right.\notag \\&\left.
    +\int_{\min\{{\epsilon_{\max,u}^{(e)}},\frac{\theta^{(e)}_{u}}{2}\}}^{\epsilon_{\max,u}^{(e)}} \mathcal{\Pr}_{\rm{cov}|\tilde{k},\tilde{\eta}}(s_{\tilde{k},\tilde{\eta},2})f_{\Delta_{u}^{(e)}}(\delta_{u}^{(e)}) \mathrm{d} \delta_{u}^{(e)}
    +\left(1-\Omega_{\delta_{u}^{(a)}}\right)\int_{\max\{\epsilon_{\min,u}^{(e)},-\frac{\theta^{(e)}_{u}}{2}\}}^{\min\{{\epsilon_{\max,u}^{(e)}},\frac{\theta^{(e)}_{u}}{2}\}} \mathcal{\Pr}_{\rm{cov}|\tilde{k},\tilde{\eta}}(s_{\tilde{k},\tilde{\eta},2})f_{\Delta_{u}^{(e)}}(\delta_{u}^{(e)}) \mathrm{d} \delta_{u}^{(e)} \right)\notag
    \\
    &+\left(1-\Omega_{\delta_{v}^{(a)}}\Omega_{\delta_{v}^{(e)}}\right)\Omega_{\delta_{u}^{(a)}}\int_{\max\{\epsilon_{\min,u}^{(e)},-\frac{\theta^{(e)}_{u}}{2}\}}^{\min\{{\epsilon_{\max,u}^{(e)}},\frac{\theta^{(e)}_{u}}{2}\}} \mathcal{\Pr}_{\rm{cov}|\tilde{k},\tilde{\eta}}(s_{\tilde{k},\tilde{\eta},3})f_{\Delta_{u}^{(e)}}(\delta_{u}^{(e)}) \mathrm{d} \delta_{u}^{(e)}
    \notag \\&
    +\left(1-\Omega_{\delta_{v}^{(a)}}\Omega_{\delta_{v}^{(e)}}\right)\left( \int_{\epsilon_{\min,u}^{(e)}}^{\max\{\epsilon_{\min,u}^{(e)},-\frac{\theta^{(e)}_{u}}{2}\}} \mathcal{\Pr}_{\rm{cov}|\tilde{k},\tilde{\eta}}(s_{\tilde{k},\tilde{\eta},4})f_{\Delta_{u}^{(e)}}(\delta_{u}^{(e)}) \mathrm{d} \delta_{u}^{(e)}\right.\notag 
    \\
    &\left.+\int_{\min\{{\epsilon_{\max,u}^{(e)}},\frac{\theta^{(e)}_{u}}{2}\}}^{\epsilon_{\max,u}^{(e)}} \mathcal{\Pr}_{\rm{cov}|\tilde{k},\tilde{\eta}}(s_{\tilde{k},\tilde{\eta},4})f_{\Delta_{u}^{(e)}}(\delta_{u}^{(e)}) \mathrm{d} \delta_{u}^{(e)}
    +\left(1-\Omega_{\delta_{u}^{(a)}}\right)\int_{\max\{\epsilon_{\min,u}^{(e)},-\frac{\theta^{(e)}_{u}}{2}\}}^{\min\{{\epsilon_{\max,u}^{(e)}},\frac{\theta^{(e)}_{u}}{2}\}} \mathcal{\Pr}_{\rm{cov}|\tilde{k},\tilde{\eta}}(s_{\tilde{k},\tilde{\eta},4})f_{\Delta_{u}^{(e)}}(\delta_{u}^{(e)}) \mathrm{d} \delta_{u}^{(e)} \right).
\end{align}
\hrulefill
\vspace*{4pt}
\end{figure*}
where $s_{\tilde{k},\tilde{\eta},\tilde{j}}=\frac{Tr_{\tilde{k},\tilde{\eta}}^{ {\alpha _{\tilde{\eta}}}}}{P_{t,\tilde{k}}A_{\tilde{\eta}}G_{\tilde{j}}}$,  $\Omega_{\delta_{\kappa}^{(\tau)}} \triangleq F_{\Delta_{\kappa}^{(\tau)}}(\theta^{(\mathrm{\tau})}_{\kappa}/2)-F_{{\Delta_{\kappa}^{(\tau)}}}(- \theta^{(\mathrm{\tau})}_{\kappa}/2)$ and $F_{\Delta_{\kappa}^{(\tau)}}(\delta_{\kappa}^{(\tau)})$ represents the CDF of $\delta_{\kappa}^{(\tau)}$.
\fi

\begin{remark}
The overall outage probability in Eq. \eqref{eq:outage_overall} studies the case of beam misalignment. It is not difficult to find that Eq. \eqref{eq:outage_overall} also covers the case of perfect alignment. Compared to the scene with misalignment, there is only the main-lobe beamforming gain $G_1$ in the typical link. Besides, the beamsteering error set $\mathbf{\delta}_{\tilde{k},0}$ no longer has any influence. As a result, by simplifying Eq. \eqref{eq:outage_overall}, here we derive the outage probability with perfect beam alignment as 
\ifCLASSOPTIONonecolumn
\begin{align}
\label{eq:P_o_per}
\begin{array}{l}
{P_{{\rm{o},\rm{p}}}}=1 - \sum_{\tilde{\eta}\in\{\LL,\NN\}}\sum_{\tilde{k} =1}^K \int_{H_{\tilde{k}}}^{\infty} \sum_{t=0}^{m_{\tilde{\eta}}-1} \frac{(-s_{\tilde{k},\tilde{\eta},1})^{t}}{t !} \frac{\mathrm{d}^{t}\exp(-s_{\tilde{k},\tilde{\eta},1}\sigma^2)\mathcal{L}_{I,\tilde{k},\tilde{\eta}}(s_{\tilde{k},\tilde{\eta},1})}{\mathrm{~d} s_{\tilde{k},\tilde{\eta},1}^{t}} {f}_{R_{\tilde{k},\tilde{\eta}}}(r_{\tilde{k},\tilde{\eta}}) \mathrm{d} r_{\tilde{k},\tilde{\eta}}  ,
\end{array}
\end{align}
\else
\begin{align}
\label{eq:P_o_per}
&{P_{{\rm{o},\rm{p}}}}=1 - \sum_{\tilde{\eta}\in\{\LL,\NN\}}\sum_{\tilde{k} =1}^K \int_{H_{\tilde{k}}}^{\infty} \sum_{t=0}^{m_{\tilde{\eta}}-1} \frac{(-s_{\tilde{k},\tilde{\eta},1})^{t}}{t !}\notag\\ &\times\frac{\mathrm{d}^{t}\exp(-s_{\tilde{k},\tilde{\eta},1}\sigma^2)\mathcal{L}_{I,\tilde{k},\tilde{\eta}}(s_{\tilde{k},\tilde{\eta},1})}{\mathrm{~d} s_{\tilde{k},\tilde{\eta},1}^{t}} {f}_{R_{\tilde{k},\tilde{\eta}}}(r_{\tilde{k},\tilde{\eta}}) \mathrm{d} r_{\tilde{k},\tilde{\eta}}  ,
\end{align}
\fi
where $s_{\tilde{k},\tilde{\eta},1}=\frac{Tr_{\tilde{k},\tilde{\eta}}^{ {\alpha _{\tilde{\eta}}}}}{P_{t,\tilde{k}}A_{\tilde{\eta}}G_{1}}$, and the conditional Laplace transform of the PDF of aggregate interference $\mathcal{L}_{I,\tilde{k},\tilde{\eta}}(s_{\tilde{k},\tilde{\eta},1})$ is almost the same as Eq. \eqref{eq:L_I} except that $\delta_{u}^{(e)}$ is set to zero. When we further set $K=1$, Eq. \eqref{eq:P_o_per} is reduced to the outage probability for the simple case considered in the literature, i.e., single tier network with perfect beam alignment.
\end{remark}
\section{Numerical Results}\label{sec:s&r}
\ifCLASSOPTIONonecolumn
\begin{small}
\begin{table}[h]
\centering
\caption{System Parameters}
\label{tab:simu para}
\begin{tabular}{|c|c|}\hline
Number of UAV tiers $K$ & $2$\\ \hline
UAV distribution heights ${H_{k}}$  & $150, 200\ \rm{m}$\\ \hline
UAV density $\lambda_k$  & $10^{-5} \ \rm{m}^{-2}$ \\ \hline
Path loss coefficient of LoS link $\alpha_L$ & $2.5$ \\ \hline
Path loss coefficient of NLoS link $\alpha_N$ & $4$ \\ \hline
Nakagami-$m$ parameter $m_L$ & $3$ \\ \hline
Nakagami-$m$ parameter $m_N$ & $1$ \\ \hline
Transmit power $P_{t,k}$ & $0\ \rm{dBW}, 2\ \rm{dBW}$ \\ \hline
Additional attenuation factor for LoS link $A_{\LL}$ & $1$ \\ \hline
Additional attenuation factor for NLoS link $A_{\NN}$ & $0.01$ \\ \hline
Noise power $\sigma ^2$ & $-130 \ \rm{dBW}$ \\ \hline
Parameter $a$ & $11.95$ \\ \hline
Parameter $b$ & $0.136$ \\ \hline
Distribution of beamsteering error $\delta_{v}^{(a)}$ and $\delta_{v}^{(e)}$ & $U( - \frac{\pi }{8},\frac{\pi }{8})$ \\ \hline
Distribution of beamsteering error $\delta_{u}^{(a)}$ and $\delta_{u}^{(e)}$ & $U( - \frac{\pi }{12},\frac{\pi }{12})$ \\ \hline
Number of antennas for the UE $N_u$& $4$\\\hline
SINR threshold $T$ & $0 \ \rm{dB}$ \\ \hline
\end{tabular}
\end{table}
\end{small}
\else
\fi
This section presents the numerical results, which provide insights into the impact of various system parameters on the overall outage performance. Monte Carlo simulation results are provided to validate the numerical results. Unless stated otherwise, the main system parameters adopted in this paper are summarized in TABLE~\ref{tab:simu para} \cite{8904396,8663615,8488493}, where the values of $a$ and $b$ correspond to the dense urban scenario. As for the main parameters of the antenna array pattern, the uniform planar square array with half-wavelength antenna element spacing is assumed at both UAVs and the typical UE. According to~\cite{venugopal2016device}, given the number of antennas $N_{\kappa}$, we have the main-lobe gain, side-lobe gain, and the half-power beamwidth in the azimuth and elevation, respectively, given by
\begin{align}
G_{\kappa}=N_{\kappa},
\end{align}
\begin{align}
g_{\kappa}=\frac{\sqrt{N_{\kappa}}-\frac{\sqrt{3}}{2 \pi} N_{\kappa} \sin \left(\frac{\sqrt{3}}{2 \sqrt{N_{\kappa}}}\right)}{\sqrt{N_{\kappa}}-\frac{\sqrt{3}}{2 \pi} \sin \left(\frac{\sqrt{3}}{2 \sqrt{N_{\kappa}}}\right)},
\end{align}
\begin{align}
\theta^{(\mathrm{a})}_{\kappa}=\theta^{(\mathrm{e})}_{\kappa}=\frac{\sqrt{3}}{\sqrt{N_{\kappa}}}.
\end{align}
\ifCLASSOPTIONonecolumn
\else
\begin{small}
\begin{table}[!h]
\centering
\caption{System Parameters}
\label{tab:simu para}
\begin{tabular}{|c|c|}\hline
Number of UAV tiers $K$ & $2$\\ \hline
UAV distribution heights ${H_{k}}$  & $150, 200\ \rm{m}$\\ \hline
UAV density $\lambda_k$  & $10^{-5} \ \rm{m}^{-2}$ \\ \hline
Path loss coefficient of LoS link $\alpha_L$ & $2.5$ \\ \hline
Path loss coefficient of NLoS link $\alpha_N$ & $4$ \\ \hline
Nakagami-$m$ parameter $m_L$ & $3$ \\ \hline
Nakagami-$m$ parameter $m_N$ & $1$ \\ \hline
Transmit power $P_{t,k}$ & $\!\!\!0\ \rm{dBW}, 2\ \rm{dBW}\!\!\!$ \\ \hline
Additional attenuation factor for LoS link $A_{\LL}$ & $1$ \\ \hline
\!\!\!\!\!\!\!\!\!\!\!\!\!\!Additional attenuation factor for NLoS link $A_{\NN}\!\!\!\!\!\!\!\!\!\!\!\!\!$ & $0.01$ \\ \hline
Noise power $\sigma ^2$ & $-130 \ \rm{dBW}$ \\ \hline
Parameter $a$ & $11.95$ \\ \hline
Parameter $b$ & $0.136$ \\ \hline
\!\!\!\!\!\!Distribution of beamsteering error $\delta_{v}^{(a)}$ and $\delta_{v}^{(e)}$ \!\!\!\!\!\!\!\!& $U( - \frac{\pi }{8},\frac{\pi }{8})$ \\ \hline
\!\!\!\!\!\!Distribution of beamsteering error $\delta_{u}^{(a)}$ and $\delta_{u}^{(e)}$\!\!\!\!\!\! & $U( - \frac{\pi }{12},\frac{\pi }{12})$ \\ \hline
Number of antennas for the UE $N_u$& $4$\\\hline
SINR threshold $T$ & $0 \ \rm{dB}$ \\ \hline
\end{tabular}
\end{table}
\end{small}
\fi

\subsection{Analysis Validation}
Fig.~\ref{fig:op_ver_ALL} plots the outage probability versus the SINR threshold under the imperfect beam alignment scenario, perfect beam alignment scenario and the scenario without beamforming, where the number of antennas at the UAV is set to be $9$ for beamforming cases. To validate our analysis, we also plot the simulation results. Fig.~\ref{fig:op_ver_ALL} shows the close match between the simulation and analytical results, which verifies the validity of our derived analytical results. 
From Fig.~\ref{fig:op_ver_ALL}, it can be seen that the outage probability for the perfect beam alignment case is far greater than the outage probability for the imperfect beam alignment case. This implies the importance of beam alignment accuracy for UAV communication. The recent works on improving the accuracy of beam alignment for UAV communication can refer to the literature, e.g., \cite{9143143,10040631}. Moreover, we can see that MAPAS performs better than CDAS in terms of the outage probability, since the closest UAV may not be able to provide the strongest signal, especially under the probabilistic channel model. We also present the impacts of $m_\textrm{L}$-value in Fig.~\ref{fig:op_ver_ALL}. It can be seen that as $m_{\LL}$ increases, the outage probability for all scenarios decreases when the SINR threshold is relatively low. However, when the SINR threshold is relatively high, we can observe the opposite trend. Generally, with the increasing of $m_{\LL}$, the fading on the transmission link becomes less severe; hence, both the power of the typical link and the interference (i.e., the numerator and denominator of SINR) increase. The interaction between these two factors leads to such trends.
\ifCLASSOPTIONonecolumn
\begin{figure}[!h]
\centering
\subfigure[Imperfect beam alignment case and the case without beamforming.]{\label{fig:op_ver_misalign_noBF}\includegraphics[width=0.7\textwidth]{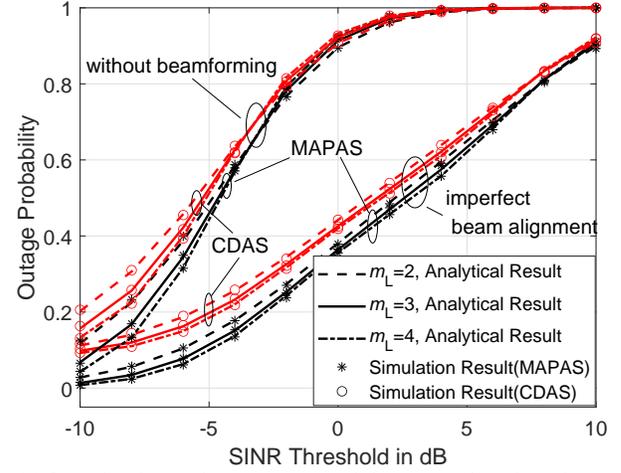}}\\
\subfigure[Perfect beam alignment case.]{\label{fig:op_ver_perf}\includegraphics[width=0.7\textwidth]{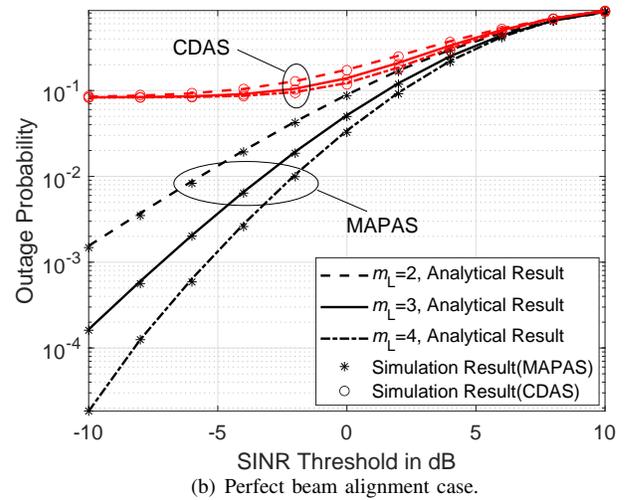}}
\caption{Outage probability versus SINR threshold for different cases.}\label{fig:op_ver_ALL}
\end{figure}
\else
\fi
\ifCLASSOPTIONonecolumn
\else
\begin{figure}[!t]
\centering
\subfigure[Imperfect beam alignment case and the case without beamforming.]{\label{fig:op_ver_misalign_noBF}\includegraphics[width=\linewidth]{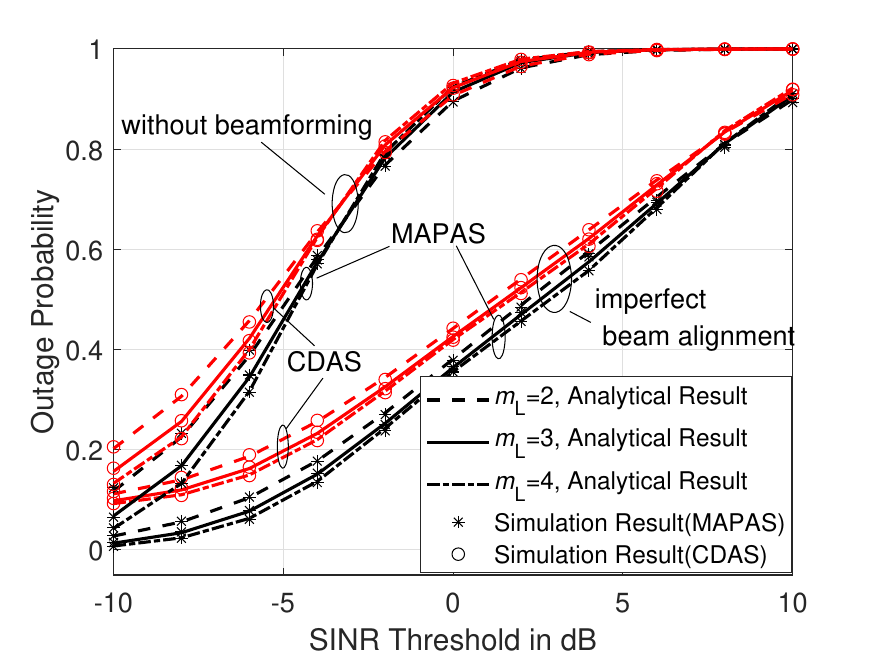}}\\
\subfigure[Perfect beam alignment case.]{\label{fig:op_ver_perf}\includegraphics[width=\linewidth]{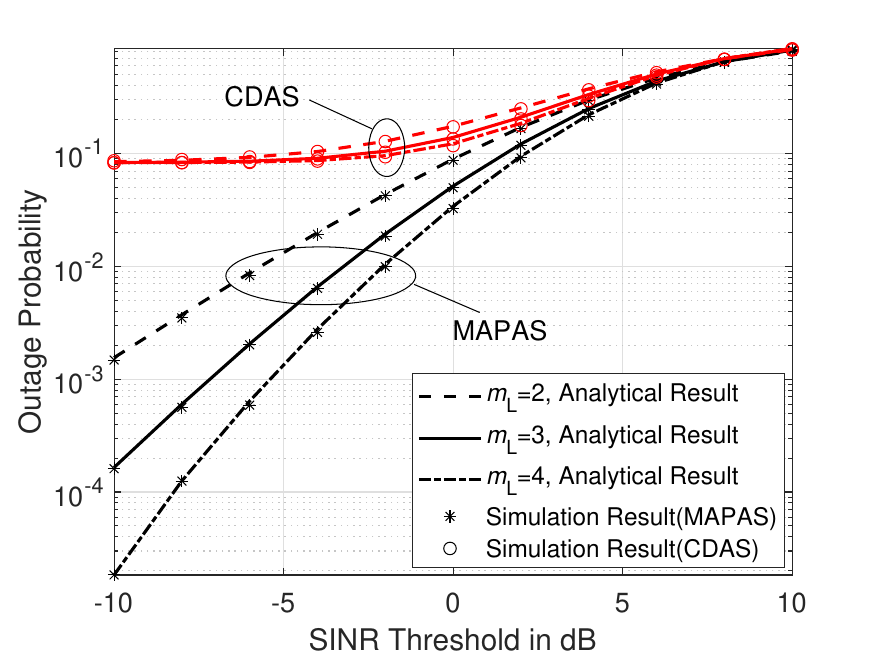}}
\caption{Outage probability versus SINR threshold for different cases.}\label{fig:op_ver_ALL}
\end{figure}
\fi

\ifCLASSOPTIONonecolumn
\begin{figure}[!h]
    \centering
        \subfigure[Imperfect beam alignment case.]{\label{fig:op_height}\includegraphics[width=0.7\textwidth]{fig/op_height.eps}}
        \subfigure[Perfect beam alignment case.]{\label{fig:op_per_height}\includegraphics[width=0.7\textwidth]{fig/op_height_perf.eps}}
      \caption{Outage probability versus the number of antennas for UAV $N_v$ under different UAV heights for both imperfect and perfect alignment cases.}\label{fig:op_number_height}  
\end{figure}
\else
\fi

\subsection{Effect of Number of Antennas}
In this section, we investigate the impact of the 3D beam pattern. Fig.~\ref{fig:op_number_height} plots the outage probability versus the number of antennas for UAVs $N_v$ under imperfect beam alignment and perfect beam alignment cases, where the height of the first tier of UAVs is respectively set to be 150 m and 200 m and the height difference between the first tier and the second tier is 50 m. The number of antennas at the UE follows TABLE~\ref{tab:simu para}. From Fig.~\ref{fig:op_height}, under the imperfect beam alignment case, it can be seen that the outage probability for both schemes first decreases and then increases as the number of antennas for UAV $N_v$ increases. This can be explained as follows. When the antenna number is in the small range, the beamwidth of the main-lobe is large, which can be even larger than the maximum beamsteering error. In other words, the beamsteering error has little impact on the outage probability. Hence, increasing the number of antennas is equivalent to the increase of main-lobe gain and the decrease of main-lobe beamwidth, which is beneficial to the SINR, thereby reducing the outage probability. However, when the number of antennas keeps increasing, the main-lobe beamwidth becomes very narrow, e.g., far smaller than the maximum beamsteering error. That is to say, the probability of losing directivity increases with the increment of the number of antennas. Consequently, as the number of antennas further increases, the outage probability becomes worse. Comparing Fig.~\ref{fig:op_height} and Fig.~\ref{fig:op_per_height}, we can observe that the difference in outage probability performance between imperfect and perfect beam alignment becomes larger as the number of transmitting antennas increases. This demonstrates that the narrower the main-lobe beamwidth, the less robust the beamforming is under the imperfect alignment case. In conclusion, under the imperfect beam alignment case, an excessive number of antennas is adverse to the network coverage probability. As for the perfect beam alignment case, Fig.~\ref{fig:op_per_height} shows that, as the number of antennas becomes large, the outage probability decreases, and the trend of decreasing gradually slows down. This figure implies that from the perspective such as hardware cost, it is not necessary to equip the UAV with too many antennas since the performance gain is very small.
\ifCLASSOPTIONonecolumn
\else
\begin{figure}[!t]
    \centering
        \subfigure[Imperfect beam alignment case.]{\label{fig:op_height}\includegraphics[width=\linewidth]{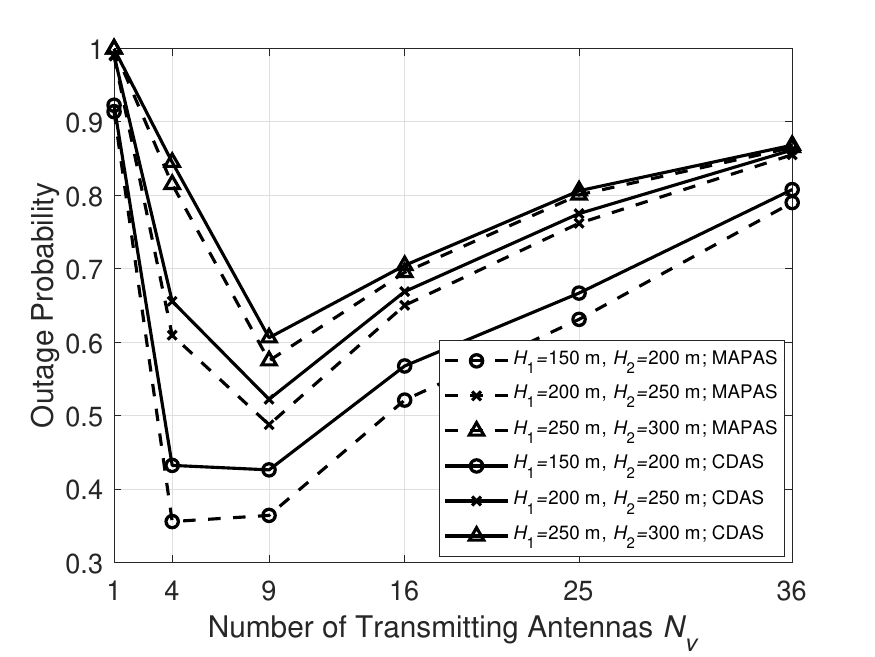}}
        \subfigure[Perfect beam alignment case.]{\label{fig:op_per_height}\includegraphics[width=\linewidth]{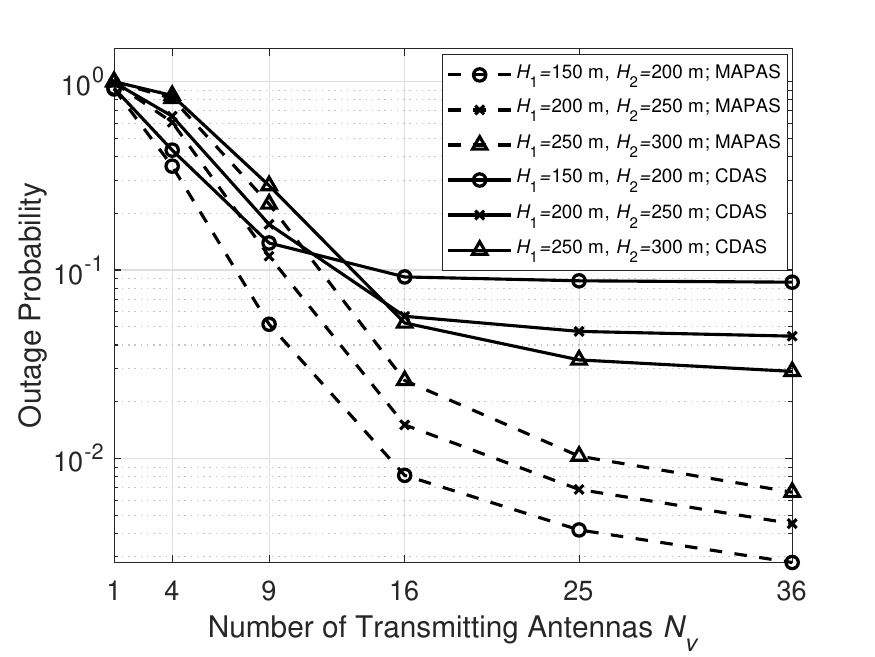}}
      \caption{Outage probability versus the number of antennas for UAV $N_v$ under different UAV heights for both imperfect and perfect alignment cases.}\label{fig:op_number_height}  
\end{figure}
\fi
\subsection{Effect of UAV Deployment Height}
With regards to the effect of UAV deployment height, Fig.~\ref{fig:op_height} shows that the higher deployment of UAVs will lead to a larger outage probability under the imperfect alignment case. The reason is as follows. The higher height implies the larger coverage area of the main-lobe beam on certain tiers, which introduces more interference from more UAVs to the typical UE with the main-lobe pointed. When the beam is mis-pointed, this kind of effect on SINR becomes worse. Besides that, for the same projection point, the higher height indicates that the transmission link is more likely to be LoS, which means the signal strength from the interfering UAV becomes stronger. Hence, when the misalignment for the beamforming is non-negligible, the lower deployment of UAVs is preferred for both schemes. Under the case of perfect beam alignment, from Fig.~\ref{fig:op_per_height}, it can be seen that the outage probability of MAPAS is smaller for the lower height of UAVs. The explanation is the same as before. However, in terms of the performance of CDAS, it seems that the outage probability for the higher altitude of UAVs can be better. This is mainly because of the fact the UE is associated with the closest UAV and their link can be either LoS or NLoS. As mentioned before, for the same projection point, the link from UAV at the lower height is much more likely to be NLoS, which reduces the desired signal strength at the UE whereby degrading the outage performance. 

\subsection{Effect of UAV Density}
Fig.~\ref{fig:op_density_2} plots the outage probability versus the number of antennas for UAV under different UAV densities for both imperfect and perfect alignment cases. Under the imperfect beam alignment scenario, as expected, the outage probability drops at first and then rises with the increase in the number of antennas. Moreover, since more interfering UAVs are involved in the system, the higher UAV density leads to worse outage probability performance. Besides that, Fig.~\ref{fig:op_density} shows that the optimal number of UAV antennas increases as the UAV density rises, e.g., for MAPAS, the optimal number antennas is 4 for $\lambda_{k}=0.5\times 10^{-5}\ \rm{m}^{-2}$ while it goes to 16 for $\lambda_{k}=5\times 10^{-5}\ \rm{m}^{-2}$. The reason is as follows. When the UAV density is very sparse, the number of interfering UAVs falling into the region covered by the main-lobe beam of the UE is very small. In other words, the interference is not that severe. Then the serving UAV needs to ensure that the typical UE is covered by its main-lobe beam; hence, a larger main-lobe beamwidth (equivalently, a smaller number of antennas) is preferred. However, when the interfering UAVs are very dense, the interference becomes very severe. One way to reduce the interference is to reduce the density of interfering UAVs with main-lobe beam pointed to the typical UE. From our analysis, this can be achieved by narrowing the main-lobe beamwidth. Note that it cannot be too narrow, because this can degrade the signal strength from the serving UAV due to the beam misalignment. Hence, a relatively larger number of antennas is preferred for the case of denser UAVs. Under the perfect beam scenario, for the MAPAS, sparse UAVs lead to a lower outage probability as expected. However, this is not the case for the CDAS. Fig.~\ref{fig:op_per_density} shows that the denser UAVs can even result in a better outage probability, especially when the number of antennas is large. The reason is as follows. When the UAV density is very sparse, the closest UAV (i.e., the serving UAV) can be very far away, which consequently leads to a very weak signal strength from the serving UAV. Increasing the density of UAVs somehow improves the signal strength from the serving UAV, which benefits SINR. However, too much interference can deteriorate the SINR performance. Hence, the interplay of these factors leads to the trend of CDAS.
\ifCLASSOPTIONonecolumn
\begin{figure}[!t]
    \centering
        \subfigure[Imperfect beam alignment case.]{\label{fig:op_density}\includegraphics[width=0.7\textwidth]{fig/op_density.eps}}
        \subfigure[Perfect beam alignment case.]{\label{fig:op_per_density}\includegraphics[width=0.7\textwidth]{fig/op_density_perf.eps}}
        \caption{Outage probability versus the number of antennas for UAV $N_v$ under different UAV densities for both imperfect and perfect alignment cases.}
        \label{fig:op_density_2}
\end{figure}
\else
\fi
\ifCLASSOPTIONonecolumn
\else
\begin{figure}[!h]
    \centering
        \subfigure[Imperfect beam alignment case.]{\label{fig:op_density}\includegraphics[width=\linewidth]{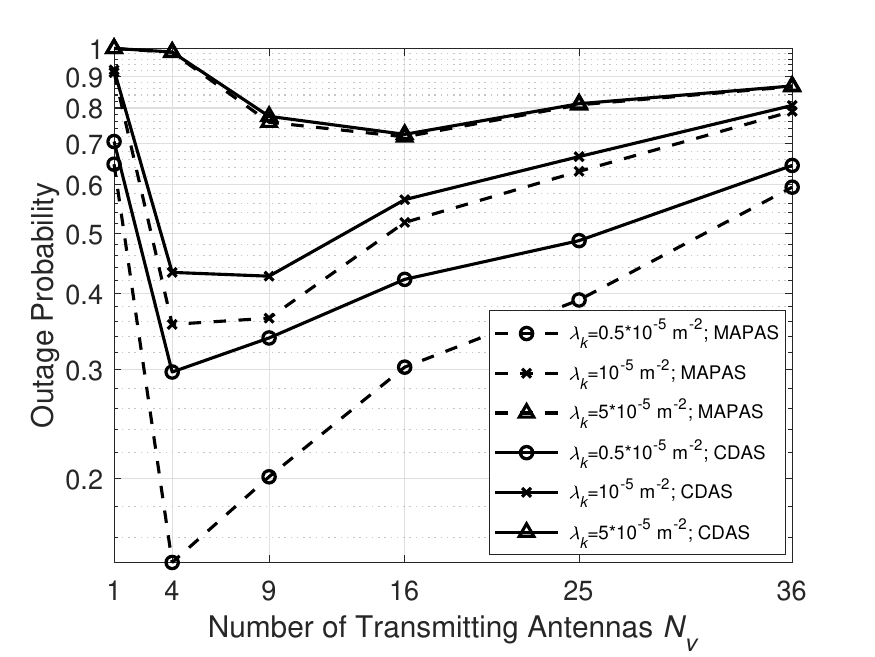}}
        \subfigure[Perfect beam alignment case.]{\label{fig:op_per_density}\includegraphics[width=\linewidth]{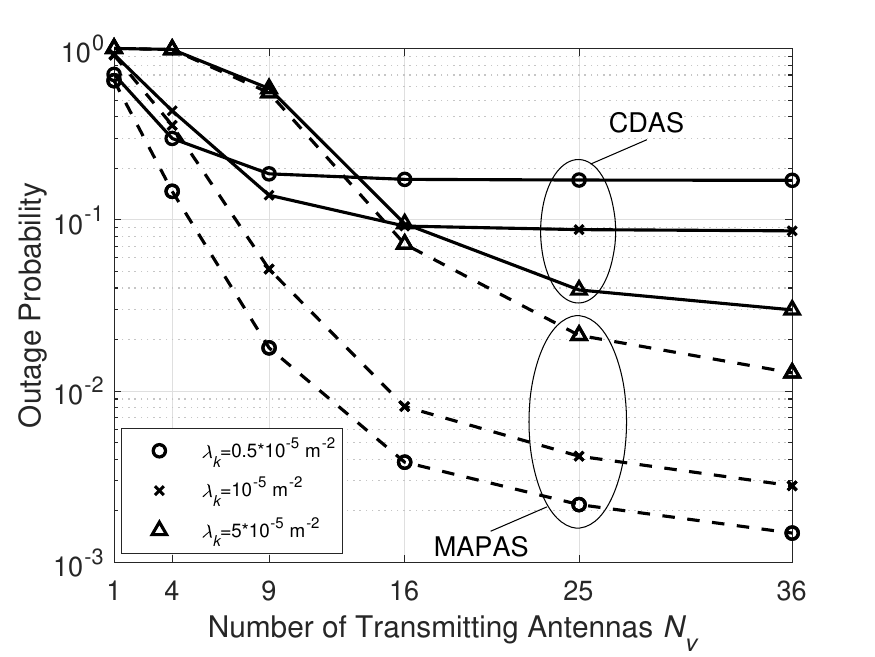}}
        \caption{Outage probability versus the number of antennas for UAV $N_v$ under different UAV densities for both imperfect and perfect alignment cases.}
        \label{fig:op_density_2}
\end{figure}
\fi
\subsection{Effect of Beamsteering Error Range}
Fig.~\ref{fig:op_misalignment} plots the outage probability versus the number of antennas for the UAV under different maximum beamsteering errors. It can be seen from this figure that a smaller beamsteering error range will lead to a lower outage probability as expected, since the typical UE is covered by the main-lobe beam in most cases when the beamsteering error range is very small. This again indicates that stable beam alignment is important to reduce interference and the outage probability. In addition, Fig.~\ref{fig:op_misalignment} reflects that the optimal number of antennas occurs when half of the main-lobe beamwidth is around the maximum beamsteering error. However, note that this is not always the case. The optimal number is also determined by other factors such as the UAV deployment height and the UAV density, as shown in Fig.~\ref{fig:op_height} and Fig.~\ref{fig:op_density}. Hence, the beamforming needs to be carefully designed for different transmission environments. 
\ifCLASSOPTIONonecolumn
\begin{figure}[htb]
    \centering
        \includegraphics[width=0.7\textwidth]{fig/op_misalignment.eps}
        \caption{Outage probability versus the number of antennas for UAV $N_v$ under different beamsteering error ranges.}
        \label{fig:op_misalignment}
\end{figure}
\else
\begin{figure}[!h]
    \centering
        \includegraphics[width=\linewidth]{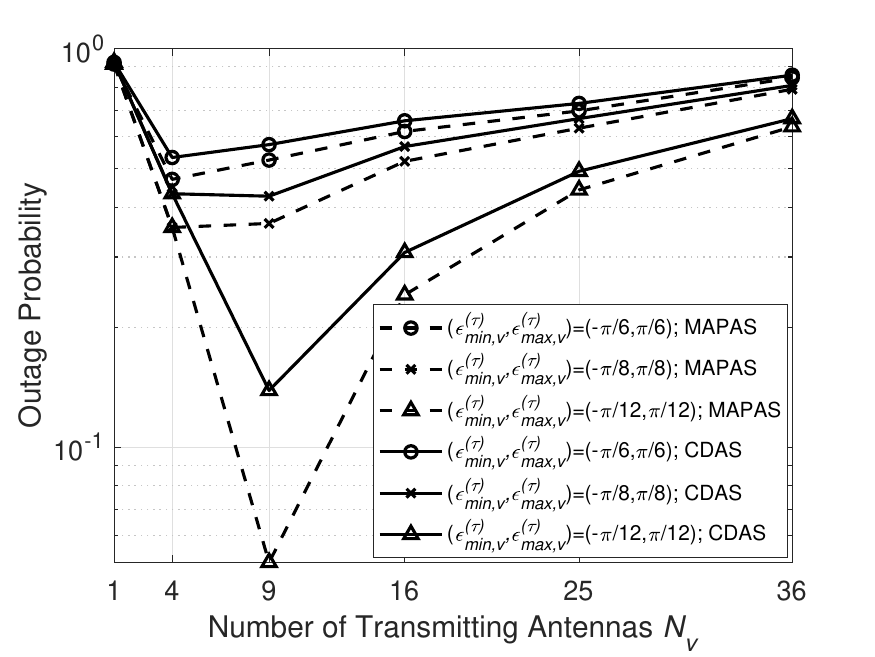}
        \caption{Outage probability versus the number of antennas for UAV $N_v$ under different beamsteering error ranges.}
        \label{fig:op_misalignment}
\end{figure}
\fi

\section{Conclusions} \label{sec:concl}
In this paper, the outage performance of UAV communication has been analyzed by taking the random beam misalignment into account. Based on stochastic geometry, a mathematical framework for computing the outage probability of the 3D multi-tier UAV communication system for two widely adopted association schemes has been developed. The analysis was validated by Monte Carlo simulations. Our results demonstrated that when beam misalignment exists, too many antennas can lose the high beamforming gain over the associated link, which degrades the outage performance. In addition, when the UAV density becomes larger, or the UAVs are deployed at a higher altitude, although the beam misalignment exists, we have to set the optimal number of antennas to be relatively larger to fight against more interference for both association schemes.

\section*{Appendix A: Proof of Lemma 1}\label{appendix:Lemma1}
\begin{IEEEproof}
 According to the geometrical relationship, the projection area formed by the main-lobe beam of the UE (in the following, it is referred to as the projection area for simplicity) on the UAV deployment tier can be represented by an isosceles trapezoid, i.e., the dark green region as shown in Fig.~\ref{fig:proj_a_new}. Let $\beta$ denote the elevation angle between the typical UE and the associated UAV on the $\tilde{k}$-th tier. When $\beta  + {\delta_{u}^{(e)}}$ is smaller than $\frac{\pi}{2}-\frac{{{\theta^{(e)}_{u}}}}{2}$, the projection region does not cover the origin of the deployment tier. Based on the geometrical relationship depicted in Fig.~\ref{fig:proj_z}, it can be calculated that the horizontal distance from the origin of the $k$-th tier to the upper line of the trapezoid is $Z_{in,\tilde{k},k}$ expressed by Eq. \eqref{eq:z_in}, the horizontal distance from the origin to the lower line is $Z_{out,\tilde{k},k}$ expressed by Eq. \eqref{eq:z_out}, and the angle between the two waists of the trapezoid is $\theta^{(\mathrm{a})}_{u}$. Note that when formulating the Laplace transform of the PDF of aggregate interference, the integration over the residing area is generally transformed from Cartesian coordinates to polar coordinates for computation simplicity. Consequently, in this work, to facilitate the analysis, the isosceles trapezoid project area is approximated by a sector area. That is to say that the isosceles trapezoid project area can be approximated to a ring sector with an angle $\theta^{(\mathrm{a})}_{u}$, an inner radius $Z_{in,\tilde{k},k}$ and an outer radius $Z_{out,\tilde{k},k}$, as shown by the dashed area in Fig.~\ref{fig:proj_a_new}. As seen from Fig.~\ref{fig:proj_a_new}, this approximation is acceptable because the two areas are almost the same, especially when the beam width is narrow.
 
 When $\beta  + {\delta_{u}^{(e)}}$ is greater than $\frac{\pi}{2}-\frac{{{\theta^{(e)}_{u}}}}{2}$, the projection range covers the origin of the deployment tier. At this time, the projection area is still an isosceles trapezoid. However, the ring sector approximated by this isosceles trapezoid is very complex, which complicates the analysis. By noticing that, due to the limitation of UAV height and density, the probability that UAV is around the origin (i.e., $\beta  + {\delta_{u}^{(e)}}$ is greater than $\frac{\pi}{2}-\frac{{{\theta^{(e)}_{u}}}}{2}$) can be very small. Hence, we continue to use the ring sector approximation mentioned before for this case. The simulation results will demonstrate the accuracy of the above approximations.
\end{IEEEproof}

\section*{Appendix B: Proof of Lemmas 2 and 3}\label{appendix:pdf}
\begin{IEEEproof}
Before deriving the PDF of serving distance when the association link is $\tilde{\eta}$-type link under the MAPAS, we need first to get the PDF of the distance $r_{k,\tilde{\eta}}$ from the typical UE to the closest UAV with $\tilde{\eta}$-type link on the $k$-th tier, which plays an important role in determining the PDF of serving distance. The locations of the $k$-th tier UAVs with $\tilde{\eta}$-type link are modeled as a HPPP. Therefore, by using the mapping theorem in stochastic geometry, we can transfer the origin HPPP to a one dimension PPP with density $2{\rm{\pi}}{\lambda_{k}}p_{k,\tilde{\eta}}(z)$. According to the void probability of the one dimension PPP, the CDF and PDF of the distance from the typical UE to the closest UAV with $\tilde{\eta}$-type link on the $k$-th tier can be respectively expressed as
\ifCLASSOPTIONonecolumn
\begin{align}
    \hat{F}_{R_{k,\tilde{\eta}}}({r_{k,\tilde{\eta}}}) &= \Pr(R_{k,\tilde{\eta}} < {r_{k,\tilde{\eta}}}) = 1 - \exp \left( { - 2{\rm{\pi }}{ \lambda_{k}}\int_0^{\sqrt {r_{k,\tilde{\eta}}^2 - {H_{k}}^2} } {z{p_{{k},\tilde{\eta}}}(z){\rm{d}}z} } \right),
\end{align}
\else
\begin{align}
    &\hat{F}_{R_{k,\tilde{\eta}}}({r_{k,\tilde{\eta}}}) = \Pr(R_{k,\tilde{\eta}} < {r_{k,\tilde{\eta}}}) \notag
    \\ &= 1 - \exp \left( { - 2{\rm{\pi }}{ \lambda_{k}}\int_0^{\sqrt {r_{k,\tilde{\eta}}^2 - {H_{k}}^2} } {z{p_{{k},\tilde{\eta}}}(z){\rm{d}}z} } \right),
\end{align}
\fi
\ifCLASSOPTIONonecolumn
and
\begin{align}
\label{eq:PDF_RN}
    {\hat{f}_{R_{k,\tilde{\eta}}}}(r_{k,\tilde{\eta}}) = \frac{{\rm{d}} \hat{F}_{R_{k,\tilde{\eta}}}({r_{k,\tilde{\eta}}})}{{\rm{d}}{r_{k,\tilde{\eta}}}} = 2{\rm{\pi }}{ \lambda_{k}}{r_{k,\tilde{\eta}}}p_{{k},\tilde{\eta}}\left(\sqrt {{r_{k,\tilde{\eta}}}^2 - {H_{k}}^2} \right)\exp \left( { - 2{\rm{\pi }}{ \lambda_{k}}\int_0^{\sqrt {{r_{k,\tilde{\eta}}}^2 - {H_{k}}^2} } {z{p_{{k},\tilde{\eta}}}(z){\rm{d}}z} } \right).
\end{align}
\else
and Eq. \eqref{eq:PDF_RN} at the top of the next page.
\begin{figure*}[!t]
\begin{align}
\label{eq:PDF_RN}
    {\hat{f}_{R_{k,\tilde{\eta}}}}(r_{k,\tilde{\eta}}) = \frac{{\rm{d}} F_{\hat{R}_{k,\tilde{\eta}}}({r_{k,\tilde{\eta}}})}{{\rm{d}}{r_{k,\tilde{\eta}}}}= 2{\rm{\pi }}{ \lambda_{k}}{r_{k,\tilde{\eta}}}p_{{k},\tilde{\eta}}\left(\sqrt {{r_{k,\tilde{\eta}}}^2 - {H_{k}}^2} \right)\exp \left( { - 2{\rm{\pi }}{ \lambda_{k}}\int_0^{\sqrt {{r_{k,\tilde{\eta}}}^2 - {H_{k}}^2} }\!\!\!\!\! {z{p_{{k},\tilde{\eta}}}(z){\rm{d}}z} } \right).
\end{align}
\end{figure*}
\fi

\ifCLASSOPTIONonecolumn
From the definition of the MAPAS, the CDF of serving distance, when the association link is the $\tilde{\eta}$-type link, can be interpreted as the probability that the average received power from the associated UAV is larger than the average received power from the other UAVs, i.e., the horizontal distance between the other UAVs and the typical UE should be greater than a specific distance, namely the equivalent distance. The formulation of the equivalent distance ${Z_{eq,k}}(r_{\tilde{k},\tilde{\eta}},\eta)$ is given in Eq. \eqref{eq:closest dis}. Mathematically, the CDF can be expressed as
\begin{align}
\label{eq:CDF_R_N}
    \begin{aligned}
        &{F}_{R_{\tilde{k},\tilde{\eta}}}(r_{\tilde{k},\tilde{\eta}})\mathop =\limits^{(a)}\Pr\left(R_{\tilde{k},\tilde{\eta}}<r_{\tilde{k},\tilde{\eta}} \cap \forall {r_{k,\eta_s} }>\sqrt{{Z^2_{eq,k}}(r_{\tilde{k},\eta_s},\tilde{\eta})+H_{k}^2}\cap \left(\forall {r_{k,\tilde{\eta}} \setminus r_{\tilde{k},\tilde{\eta}}}\right)>\sqrt{{Z^2_{eq,k}}(r_{\tilde{k},\tilde{\eta}},\tilde{\eta})+H_{k}^2}\right) \\
    &=\int_{H_{\tilde{k}}}^{r_{\tilde{k},\tilde{\eta}}} \exp \left(-2 \pi \!\!\!\!\!\sum\limits_{k \in\{1,2, \ldots K\}} \!\!\!\!\!\lambda_{k} \int_{0}^{{Z_{eq,k}}(r_{\tilde{k},\tilde{\eta}},\eta_s)} \!\!\!\!\!\!\!\!\!\!p_{k,\eta_s}(z) z \mathrm{~d} z -2 \pi\!\!\!\!\! \!\!\!\sum\limits_{k \in\{1,2, \ldots, K\} \setminus \tilde{k}}\!\!\!\!\! \lambda_{k} \int_{0}^{{Z_{eq,k}}(r_{\tilde{k},\tilde{\eta}},\tilde{\eta})}\!\!\!\!\!\!\!\!\!\!p_{k,\tilde{\eta}}(z) z \mathrm{~d} z\right) \hat{f}_{R_{\tilde{k},\tilde{\eta}}}(r_{\tilde{k},\tilde{\eta}}) \mathrm{d} r_{\tilde{k},\tilde{\eta}}.
    \end{aligned}
\end{align}
\else
From the definition of the MAPAS, the CDF of serving distance, when the association link is the $\tilde{\eta}$-type link, can be interpreted as the probability that the average received power from the associated UAV is larger than the average received power from the other UAVs, i.e., the horizontal distance between the other UAVs and the typical UE should be greater than a specific distance, namely the equivalent distance. The formulation of the equivalent distance ${Z_{eq,k}}(r_{\tilde{k},\tilde{\eta}},\eta)$ is given in Eq. \eqref{eq:closest dis}. Mathematically, the CDF can be expressed as Eq. \eqref{eq:CDF_R_N} at the top of the next page,
\begin{figure*}
\begin{align}
\label{eq:CDF_R_N}
    \begin{aligned}
    &{F}_{R_{\tilde{k},\tilde{\eta}}}(r_{\tilde{k},\tilde{\eta}})\mathop =\limits^{(a)}\Pr\left(R_{\tilde{k},\tilde{\eta}}<r_{\tilde{k},\tilde{\eta}} \cap \forall {r_{k,\eta_s} }>\sqrt{{Z^2_{eq,k}}(r_{\tilde{k},\eta_s},\tilde{\eta})+H_{k}^2}\cap \left(\forall {r_{k,\tilde{\eta}} \setminus r_{\tilde{k},\tilde{\eta}}}\right)>\sqrt{{Z^2_{eq,k}}(r_{\tilde{k},\tilde{\eta}},\tilde{\eta})+H_{k}^2}\right) \\
    &=\int_{H_{\tilde{k}}}^{r_{\tilde{k},\tilde{\eta}}} \exp \left(-2 \pi \!\!\!\!\!\sum\limits_{k \in\{1,2, \ldots K\}} \!\!\!\!\!\lambda_{k} \int_{0}^{{Z_{eq,k}}(r_{\tilde{k},\tilde{\eta}},\eta_s)} \!\!\!\!\!\!\!\!\!\!p_{k,\eta_s}(z) z \mathrm{~d} z -2 \pi\!\!\!\!\! \!\!\!\sum\limits_{k \in\{1,2, \ldots, K\} \setminus \tilde{k}}\!\!\!\!\! \lambda_{k} \int_{0}^{{Z_{eq,k}}(r_{\tilde{k},\tilde{\eta}},\tilde{\eta})}\!\!\!\!\!\!\!\!\!\!p_{k,\tilde{\eta}}(z) z \mathrm{~d} z\right) \hat{f}_{R_{\tilde{k},\tilde{\eta}}}(r_{\tilde{k},\tilde{\eta}}) \mathrm{d} r_{\tilde{k},\tilde{\eta}}.
    \end{aligned}
\end{align}
\end{figure*}
\fi
\ifCLASSOPTIONonecolumn
where $(a)$ ensures that the average received power from the associated UAV is larger than the average received power from the other UAVs. Then we can derive the PDF of the serving distance when the association link is the $\tilde{\eta}$-type link as shown in Eq. \eqref{eq:PDF_R_N_result} by taking the derivative of Eq. \eqref{eq:CDF_R_N} with respect to $r_{\tilde{k},\tilde{\eta}}$.
\else
where $(a)$ ensures that the average received power from the associated UAV is larger than the average received power from the other UAVs. Then we can derive the PDF of the serving distance when the association link is the $\tilde{\eta}$-type link as shown in Eq. \eqref{eq:PDF_R_N_result} by taking the derivative of Eq. \eqref{eq:CDF_R_N} with respect to $r_{\tilde{k},\tilde{\eta}}$.
\fi

For the CDAS, the UE is always associated with the closest UAV. Applying the void probability of the PPP, we can get the PDF of the closest distance as
\begin{align}
\label{eq:PDF_CD}
\begin{aligned}
\hat{f}_{R_{\tilde{k},\tilde{\eta}}}(r_{\tilde{k},\tilde{\eta}})=2{\rm{\pi }}{\lambda_{\tilde{k}}}{r_{\tilde{k},\tilde{\eta}}} \exp \!\!\left(\!\!- \pi \!\!\!\!\!\!\sum_{{k} \in\{1,2, \ldots K\}}\!\!\!\!\! \lambda_{k}{Z^2_{eq,k}}(r_{\tilde{k},\tilde{\eta}})\!\!\right).
    \end{aligned}
\end{align}

Multiplying Eq. \eqref{eq:PDF_CD} by the probability that the serving UAV is in $\tilde{\eta}$-type link, we can get the PDF of serving distance under the CDAS denoted by Eq. \eqref{eq:PDF_CDAS}. 

\end{IEEEproof}

\section*{Appendix C: Proof of Proposition 1}\label{appendix:LT}

\begin{IEEEproof}
\ifCLASSOPTIONonecolumn
    According to the definition of Laplace transform, Eq. \eqref{eq:L_I} can be further expressed as
\begin{align}
    \mathcal{L}_{I,\tilde{k},\tilde{\eta}}(s)=&\mathbb{E}_{{\Phi_{k,j,\eta}},h_{\eta,j,i}}\left[\exp \left(-s\sum\limits_{k=1}^{K}\sum\limits_{j=1}^{4}\sum\limits_{\eta \in {\LL,\NN}}\sum\limits_{x_{k,i} \in \Phi_{k,j,\eta}\setminus x_{\tilde{k},0}}  A_{\eta} P_{\mathrm{t},k} M_{k,j,i} h_{\eta,j,i} \|x_{k,i}\|^{-\alpha_{\eta}}\right)\right]\notag \\
    {\mathop =\limits^{(a)}}& \prod\limits_{k=1}^{K}\prod\limits_{j=1}^{4}\prod\limits_{\eta \in {\LL,\NN}} \mathbb{E}_{{\Phi_{k,j,\eta}},h_{\eta,j,i}}\left[\prod\limits_{{x_{k,i}} \in {\Phi_{k,j,\eta}/ x_{\tilde{k},0}}} \exp \left(-s A_{\eta} P_{\mathrm{t},k} M_{k,j,i} h_{\eta,j,i} {\|x_{k,i}\|}^{-\alpha_{\eta}}\right)\right],
\end{align}
where step $(a)$ is derived by the independence between PPPs.

The next step is to find the exact formula for the expectation terms in Eq. \eqref{eq:L_I}. Taking $j=1$ as an example and we have
\begin{align}
    &\left.\mathbb{E}_{{\Phi_{k,j,\eta}},h_{\eta,j,i}}\left[\prod\limits_{{x_{k,i}} \in {\Phi_{k,j,\eta}\setminus x_{\tilde{k},0}}} \exp \left(-s A_{\eta} P_{\mathrm{t},k} M_{k,j,i} h_{\eta,j,i} {\|x_{k,i}\|}^{-\alpha_{\eta}}\right)\right]\right|_{j=1}\notag\\
    =&\mathbb{E}_{{\Phi_{k,1,\eta}}}\left[\prod\limits_{{x_{k,i}} \in {\Phi_{k,1,\eta}\setminus x_{\tilde{k},0}}} \mathbb{E}_{h_{\eta,1,i}}\left(\exp \left(-s A_{\eta} P_{\mathrm{t},k} M_{k,1,i} h_{\eta,1,i} {\|x_{k,i}\|}^{-\alpha_{\eta}}\right)\right)\right]\notag\\
    {\mathop  = \limits^{(b)}}& \mathbb{E}_{{\Phi_{k,1,\eta}}}\left[\prod\limits_{{x_{k,i}} \in {\Phi_{k,1,\eta}\setminus x_{\tilde{k},0}}} \left({\frac{{{m_{\eta}}}}{{{m_{\eta}} + s{A_{\eta}}{P_{t,k}}{M_{k,1,i}}{\|x_{k,i}\|}^{ - \alpha_{\eta} }}}} \right)^{{m_{\eta}}} \right]\notag\\
    {\mathop  = \limits^{(c)}} &\exp\left[ { - \int_{ \mathcal{S}_k\left(r_{\tilde{k},\tilde{\eta}},\delta_{u}^{(e)},\theta^{(\mathrm{a})}_{u},\theta^{(\mathrm{e})}_{u}\right)} \left(1 - \mathcal{M}\left(\eta,k,1,i,z\right)\right) {p_{k,\eta}}(z)\lambda_{k,1}{\rm{d}}z} \right]\notag\\
    {\mathop  = \limits^{(d)}} &\exp\left[ { - \theta^{(\mathrm{a})}_{u}\lambda_{k,1} \int_{\max \left\{{Z_{in,\tilde{k},k}},{{Z_{eq,k}}(r_{\tilde{k},\tilde{\eta}},\eta)}\right\}}^{\max \left\{{Z_{out,\tilde{k},k}},{{Z_{eq,k}}(r_{\tilde{k},\tilde{\eta}},\eta)}\right\}} \left(1 - \mathcal{M}\left(\eta,k,1,i,z\right)\right){p_{k,\eta}}(z)z{\rm{d}}z} \right],
\end{align}
where step $(b)$ is derived from the moment generating function of Gamma distribution.
The PGFL of PPP $\Phi_{k,1,\eta}/x_{\tilde{k},0}$ yields step $(c)$. As for the limit of the integration specified in step $(d)$, on the one hand, we need to consider the residing region of UAVs with beamforming gain $M_{k,1,i}$, which is listed in TABLE~\ref{tab:align}. On the other hand, the distance requirement for the interfering UAVs for a certain association scheme needs to be taken into account. The derivations of the other $j$ values are similar and are omitted here. 
Thus, we can deduce the results presented in Proposition~\ref{prop:L_I}.
\else
    According to the definition of Laplace transform, Eq. \eqref{eq:L_I} can be further expressed as Eq. \eqref{eq:L_I_proof} at the top of the next page,
\begin{figure*}[!t]
\begin{align}
\label{eq:L_I_proof}
    \mathcal{L}_{I,\tilde{k},\tilde{\eta}}(s)=&\mathbb{E}_{{\Phi_{k,j,\eta}},h_{\eta,j,i}}\left[\exp \left(-s\sum\limits_{k=1}^{K}\sum\limits_{j=1}^{4}\sum\limits_{\eta \in {\LL,\NN}}\sum\limits_{x_{k,i} \in \Phi_{k,j,\eta}\setminus x_{\tilde{k},0}}  A_{\eta} P_{\mathrm{t},k} M_{k,j,i} h_{\eta,j,i} \|x_{k,i}\|^{-\alpha_{\eta}}\right)\right]\notag \\
    {\mathop =\limits^{(a)}}& \prod\limits_{k=1}^{K}\prod\limits_{j=1}^{4}\prod\limits_{\eta \in {\LL,\NN}} \mathbb{E}_{{\Phi_{k,j,\eta}},h_{\eta,j,i}}\left[\prod\limits_{{x_{k,i}} \in {\Phi_{k,j,\eta}/ x_{\tilde{k},0}}} \exp \left(-s A_{\eta} P_{\mathrm{t},k} M_{k,j,i} h_{\eta,j,i} {\|x_{k,i}\|}^{-\alpha_{\eta}}\right)\right].
\end{align}
\end{figure*}
where step $(a)$ is derived by the independence between PPPs.
\begin{figure*}[!t]
\begin{align}
\label{eq:L_I_sub_proof}
    &\left.\mathbb{E}_{{\Phi_{k,j,\eta}},h_{\eta,j,i}}\left[\prod\limits_{{x_{k,i}} \in {\Phi_{k,j,\eta}\setminus x_{\tilde{k},0}}} \exp \left(-s A_{\eta} P_{\mathrm{t},k} M_{k,j,i} h_{\eta,j,i} {\|x_{k,i}\|}^{-\alpha_{\eta}}\right)\right]\right|_{j=1}\notag\\
    =&\mathbb{E}_{{\Phi_{k,1,\eta}}}\left[\prod\limits_{{x_{k,i}} \in {\Phi_{k,1,\eta}\setminus x_{\tilde{k},0}}} \mathbb{E}_{h_{\eta,1,i}}\left(\exp \left(-s A_{\eta} P_{\mathrm{t},k} M_{k,1,i} h_{\eta,1,i} {\|x_{k,i}\|}^{-\alpha_{\eta}}\right)\right)\right]\notag\\
    {\mathop  = \limits^{(b)}}& \mathbb{E}_{{\Phi_{k,1,\eta}}}\left[\prod\limits_{{x_{k,i}} \in {\Phi_{k,1,\eta}\setminus x_{\tilde{k},0}}} \left({\frac{{{m_{\eta}}}}{{{m_{\eta}} + s{A_{\eta}}{P_{t,k}}{M_{k,1,i}}{\|x_{k,i}\|}^{ - \alpha_{\eta} }}}} \right)^{{m_{\eta}}} \right]\notag\\
    {\mathop  = \limits^{(c)}} &\exp\left[ { - \int_{ \mathcal{S}_k\left(r_{\tilde{k},\tilde{\eta}},\delta_{u}^{(e)},\theta^{(\mathrm{a})}_{u},\theta^{(\mathrm{e})}_{u}\right)} \left(1 - \mathcal{M}\left(\eta,k,1,i,z\right)\right) {p_{k,\eta}}(z)\lambda_{k,1}{\rm{d}}z} \right]\notag\\
    {\mathop  = \limits^{(d)}} &\exp\left[ { - \theta^{(\mathrm{a})}_{u}\lambda_{k,1} \int_{\max \left\{{Z_{in,\tilde{k},k}},{{Z_{eq,k}}(r_{\tilde{k},\tilde{\eta}},\eta)}\right\}}^{\max \left\{{Z_{out,\tilde{k},k}},{{Z_{eq,k}}(r_{\tilde{k},\tilde{\eta}},\eta)}\right\}} \left(1 - \mathcal{M}\left(\eta,k,1,i,z\right)\right){p_{k,\eta}}(z)z{\rm{d}}z} \right].
\end{align}
\hrulefill
\end{figure*}
The next step is to find the exact formula for the expectation terms in Eq. \eqref{eq:L_I}. Taking $j=1$ as an example and we have Eq. \eqref{eq:L_I_sub_proof} at the top of the next page,
where step $(b)$ is derived from the moment generating function of Gamma distribution.
The PGFL of PPP $\Phi_{k,1,\eta}/x_{\tilde{k},0}$ yields step $(c)$. As for the limit of the integration specified in step $(d)$, on the one hand, we need to consider the residing region of UAVs with beamforming gain $M_{k,1,i}$, which is listed in TABLE~\ref{tab:align}. On the other hand, the distance requirement for the interfering UAVs for certain association scheme needs to be taken into account. The derivations of the other $j$ values are similar and are omitted here. 
Thus, we can deduce the results presented in Proposition~\ref{prop:L_I}.
\fi
\end{IEEEproof}

 \bibstyle{IEEEtran}
   \bibliography{IEEEabrv,UAV_beam_ref}

\end{document}